\pdfoutput=1
\documentclass[a4paper,onecolumn,unpublished,nopdfoutputerror]{quantumarticle}
\usepackage{amsmath,amssymb,amsthm,mathtools}
\IfFileExists{dsfont.sty}{\usepackage{dsfont}\newcommand{\id}{\mathds{1}}}{\newcommand{\id}{\mathbb{1}}}
\usepackage{braket}
\usepackage{booktabs,array}
\usepackage{tikz}
\usetikzlibrary{arrows.meta,decorations.pathreplacing,decorations.pathmorphing,shapes.geometric}
\usepackage{url}
\newtheorem{algorithm}{Algorithm}
\usepackage[numbers,sort&compress]{natbib}
\usepackage{hyperref}
\hypersetup{hidelinks,pdftitle={On the estimation of linear-optical amplitudes and probabilities with entangled input states},pdfauthor={Jean Senellart, Pierre-Emmanuel Emeriau},pdfsubject={additive-error estimation, linear optics, matrix-product states},pdfkeywords={boson sampling, Gurvits, matrix product states, quantum emitters, photonic machine learning}}
\IfFileExists{nowidow.sty}{\usepackage[all]{nowidow}}{}

\DeclareMathOperator{\per}{Per}
\DeclareMathOperator{\Tr}{Tr}
\newcommand{\C}{\mathbb C}
\newcommand{\E}{\mathbb E}
\newcommand{\bx}{\mathbf x}
\newcommand{\bt}{\mathbf t}
\newcommand{\bs}{\mathbf s}

\newcommand{\mref}[1]{\ref{#1}}
\newcommand{\meqref}[1]{\eqref{#1}}
\makeatletter
\patchcmd{\@titleatfontsize}{\@title}{\@title\par}{}{}
\makeatother

\newtheorem{theorem}{Theorem}
\newtheorem{corollary}{Corollary}
\newtheorem{proposition}{Proposition}

\theoremstyle{definition}

\newtheorem{definition}{Definition}
\theoremstyle{remark}
\newtheorem{remark}{Remark}

\title{On the estimation of linear-optical amplitudes and probabilities with entangled input states}
\author{Jean Senellart}
\affiliation{Quandela, 7 Rue L\'eonard de Vinci, 91300 Massy, France}
\author{Pierre-Emmanuel Emeriau}
\affiliation{Quandela, 7 Rue L\'eonard de Vinci, 91300 Massy, France}

\begin{document}
\sloppy
\maketitle

\begin{abstract}
Most proposed applications of boson samplers, including photonic machine-learning models, do not use the samples themselves but estimates of output probabilities, marginals or overlaps, obtained to additive precision from a finite number of shots. For product inputs such as single photons, these quantities can already be estimated classically with Gurvits's algorithm and its generalisations. We ask whether entangled inputs change this picture. For any $n$-photon input that is a matrix-product state of bond dimension $\chi$, with one photon per site and $L$ modes per site, we give an unbiased estimator of any output amplitude of an $m$-mode interferometer, whose mean-square error after $N$ draws is at most $\|\psi\|^2/N$ whatever the interferometer, the output pattern and the entanglement of the state. Each draw costs $O(nL\chi^2+nL^3+mn)$ operations after $O(nL\chi^3)$ preprocessing, and an additive error $\varepsilon$ with probability at least $1-\delta$ requires $O(\|\psi\|^2\varepsilon^{-2}\log(1/\delta))$ draws, with memory of the order of the input size. The main application is to light emitted sequentially by quantum emitters, such as spin-based sources of photonic cluster states, for which $\chi$ is bounded by the dimension of the source's retained quantum memory: one spin gives $\chi=2$ at any length, and a memory of $k$ qubits, held in emitters or in photons stored in a delay line, gives $\chi\le2^k$, so the cost stays polynomial as long as $k=O(\log n)$. The same sampler also estimates traced-memory probabilities, overlaps and individual marginal probabilities. All guarantees are additive: the estimator gives neither relative estimates of exponentially small probabilities nor a sampler of the output distribution.
\end{abstract}

\section{Introduction}
\label{sec:intro}

Linear optics with single photons is one of the most studied routes for quantum computational advantage. In boson sampling, $n$ indistinguishable photons enter an $m$-mode interferometer and are detected at its output. Each output probability is the squared modulus of the permanent of an $n\times n$ submatrix of the interferometer, and Aaronson and Arkhipov showed that sampling exactly from this distribution is not possible efficiently on a classical computer unless the polynomial hierarchy collapses, and that the same holds for approximate sampling under two conjectures about permanents of Gaussian matrices \cite{AA}. The best known exact classical samplers take time of order $n2^n$ \cite{CliffordClifford}. The first experiments followed within two years \cite{Broome,Spring}, single-photon experiments have since reached $20$ photons in $60$ modes \cite{Wang2019}, and a Gaussian variant of the problem \cite{Hamilton} was used in the first photonic claims of quantum computational advantage \cite{Zhong,Madsen}.

Sampling is a hard task to turn into an application. Much recent work therefore uses boson samplers as parametrised models, for example for supervised learning or generative modelling \cite{Gan,Sedrakyan,Gottlieb}. These applications rarely consume raw samples. The loss function, its gradient or a kernel entry is built from output probabilities, marginals or expectation values, which the device estimates from a finite number of shots. Their precision is therefore additive, of order $1/\sqrt N$ after $N$ shots, and this is precisely the regime in which classical algorithms are strong. Gurvits gave a randomised algorithm that estimates the permanent of a matrix of bounded norm to additive error $\varepsilon$ with $O(1/\varepsilon^2)$ draws \cite{Gurvits}; Aaronson and Hance extended it to outputs with repeated modes \cite{AH}; Lim and Oh recast it as an estimator over coherent states, whose variance bound does not depend on the input state, and made it efficient for product inputs \cite{LimOh}; and Seron et al.\ estimated linear statistics of boson sampling in the same spirit \cite{Seron}. For the inputs used in practice these tools make the loss functions classically estimable. This has a useful side, since models can be trained on a classical computer and only deployed on the hardware \cite{RecioArmengol,Gottlieb}. It also means that a photonic learning model whose value rests only on such estimates leaves no room for a quantum advantage, a conclusion supported by recent results on concentration in passive linear optics \cite{MonbroussouConc}.

There are two natural ways out. The first is to make the circuit itself nonlinear. Non-linear boson sampling adds photon--photon interactions inside the interferometer and raises the complexity of the task \cite{Spagnolo}. Deterministic interactions between single photons are, however, difficult to engineer: current demonstrations use a single quantum emitter coupled to a nanophotonic waveguide \cite{LeJeannic}, and placing many such elements inside a large interferometer would add substantial loss. Adaptive measurements with feed-forward provide another kind of nonlinearity \cite{ChabaudMarkhamSohbi,Monbroussou}. 
The second way is to move the nonlinearity to the source. A quantum emitter, such as a semiconductor spin or a trapped atom, can stay entangled with the photons it has already emitted and pass these correlations on to the next photon. In this way a single spin produces long linear cluster states \cite{LindnerRudolph}, as demonstrated with quantum dots and atoms \cite{Cogan,Thomas,Coste,Huet}. Spin--photon entanglement is also the basic resource of spin-optical architectures, where photons are used to entangle distant spins rather than to carry a long entangled state themselves \cite{SPOQC}.

This raises a simple question: does entanglement at the input make the additive estimation tasks hard again? The existing algorithms do not answer it. 
The algorithm of Lim and Oh is efficient for inputs that are product states across modes. It relies on sampling the Husimi function of the input, which for such states factorises into independent single-mode distributions that can be drawn one mode at a time. For an entangled input this distribution no longer factorises, and it is not clear a priori that it can be sampled efficiently.

\begin{figure}[t]
\centering
\begin{tikzpicture}[
  font=\small,
  >={Stealth[length=2mm]},
  tensor/.style={circle,draw=black!80,fill=blue!10,line width=0.6pt,minimum size=9.5mm,inner sep=0pt,font=\footnotesize},
  ctensor/.style={circle,draw=black!80,fill=orange!12,line width=0.6pt,minimum size=9.5mm,inner sep=0pt,font=\footnotesize},
  bvec/.style={regular polygon,regular polygon sides=3,draw=black!80,fill=black!8,line width=0.6pt,minimum size=6.5mm,inner sep=0pt},
  bond/.style={line width=1.1pt,black!75},
  phys/.style={line width=0.6pt,black!75},
  mode/.style={line width=0.5pt,black!70},
  env/.style={rounded corners=2pt,draw=black!80,fill=black!6,line width=0.6pt},
  past/.style={rounded corners=2pt,draw=black!80,fill=blue!5,line width=0.6pt,font=\footnotesize},
]

\node[anchor=west,font=\bfseries] at (-0.75,2.05) {(a)};

\draw[line width=0.7pt,fill=blue!6,rounded corners=3pt] (-0.55,-1.05) rectangle (0.55,1.05);
\foreach \y in {0.7,0.2,-0.72}{
  \fill[blue!18] (0.05,\y) circle (0.19);
  \draw[->,line width=0.9pt,blue!55!black] (0.05,\y-0.16) -- (0.05,\y+0.18);
}
\node[font=\footnotesize] at (0.05,-0.22) {$\vdots$};
\draw[decorate,decoration={snake,amplitude=0.6pt,segment length=3pt},black!50,line width=0.4pt] (-0.33,0.7) -- (-0.33,-0.72);
\node[align=center,font=\footnotesize] at (0.3,1.42) {$k$ emitters, memory $\C^{\chi}$, $\chi\le2^k$};

\draw[->,black!60,line width=0.5pt] (0,-1.05) -- (0,-1.45);
\draw[line width=0.6pt,fill=white,rounded corners=1pt] (-0.3,-1.85) rectangle (0.3,-1.45);
\draw[line width=0.5pt] (-0.2,-1.78) arc (160:20:0.21);
\draw[->,line width=0.5pt] (0,-1.8) -- (0.14,-1.55);
\node[anchor=east,font=\footnotesize,align=right] at (-0.4,-1.65) {memory\\[-1pt]readout};
\draw[dotted,black!55,line width=0.6pt] (0.3,-1.75) to[out=-20,in=150] (0.55,-2.6);

\draw[mode] (0.55,0) -- (6.9,0);
\draw[->,black!60] (6.3,0.3) -- (6.8,0.3);

\foreach \px/\lab in {1.8/n,4.3/2,5.75/1}{
  \draw[red!70!black,line width=0.8pt,domain=-0.34:0.34,samples=80,smooth]
    plot ({\px+\x},{0.27*exp(-(\x/0.14)^2)*cos(deg(\x*40))});
  \node[font=\footnotesize] at (\px,0.55) {$x_{\lab}$};
}
\node at (3.05,0) [fill=white,inner sep=1pt] {$\cdots$};
\draw[decorate,decoration={brace,amplitude=4pt,mirror},line width=0.5pt] (1.4,-0.45) -- (6.1,-0.45)
  node[midway,below=4pt,font=\footnotesize] {$n$ photons, one per site, $L$ modes per site};

\draw[line width=0.7pt,fill=black!4,rounded corners=2pt] (6.9,-1.62) rectangle (8.5,1.62);
\node[font=\Large] at (7.7,0) {$U$};
\node[font=\footnotesize] at (7.7,-1.92) {$m$ modes};
\foreach \i in {0,...,7}{
  \pgfmathsetmacro{\y}{1.4-\i*0.4}
  \draw[mode] (8.5,\y) -- (9.0,\y);
  \draw[line width=0.6pt,fill=black!70] (9.0,\y-0.13) -- (9.0,\y+0.13) arc (90:-90:0.13) -- cycle;
}
\draw[decorate,decoration={brace,amplitude=4pt},line width=0.5pt] (9.35,1.55) -- (9.35,-1.55)
  node[midway,right=5pt,align=left,font=\footnotesize] {output\\[-1pt]pattern $\mathbf t$};

\begin{scope}[yshift=-2.75cm]
  \node[bvec,shape border rotate=90] (f) at (0.75,0) {};
  \node[font=\footnotesize] at (0.75,-0.6) {$\bra f$};
  \node[tensor] (Mn) at (1.8,0) {$M^{(n)}$};
  \node (dots) at (3.05,0) {$\cdots$};
  \node[tensor] (M2) at (4.3,0) {$M^{(2)}$};
  \node[tensor] (M1) at (5.75,0) {$M^{(1)}$};
  \node[bvec,shape border rotate=-90] (s0) at (6.75,0) {};
  \node[font=\footnotesize] at (6.75,-0.6) {$\ket{s_0}$};
  \draw[bond] (f) -- (Mn) -- (dots) -- (M2) -- (M1) -- (s0);
  \foreach \a/\b in {Mn/dots,M2/M1}{
    \path (\a) -- node[above=0pt,font=\scriptsize] {$\chi$} (\b);
  }
  \foreach \n/\lab in {Mn/n,M2/2,M1/1}{
    \draw[phys,dashed,black!45] (\n.north) -- ++(0,1.25);
  }
  \node[anchor=west,font=\footnotesize] at (7.35,0) {$\psi(\mathbf x)=\bra f\,M^{(n)}_{x_n}\cdots M^{(1)}_{x_1}\ket{s_0}$};
\end{scope}

\begin{scope}[yshift=-6.2cm]
  \node[anchor=west,font=\bfseries] at (-0.75,2.0) {(b)};
  \node[env,minimum width=20mm,minimum height=19mm] (R) at (1.0,0) {};
  \node at (1.0,0) {$R_k$};
  \node[tensor] (Mk) at (3.2,0.6) {$M^{(k)}$};
  \node[ctensor] (Mkc) at (3.2,-0.6) {$\bar M^{(k)}$};
  \node[past,minimum width=11mm,minimum height=7mm] (a) at (5.2,0.6) {$a_{k-1}$};
  \node[past,minimum width=11mm,minimum height=7mm] (ac) at (5.2,-0.6) {$\bar a_{k-1}$};
  \draw[bond] (R.east|-Mk) -- (Mk) -- (a);
  \draw[bond] (R.east|-Mkc) -- (Mkc) -- (ac);
  \draw[phys] (Mk.north) -- ++(0,0.4) node[right,font=\footnotesize] {$v_k$};
  \draw[phys] (Mkc.south) -- ++(0,-0.4) node[right,font=\footnotesize] {$\bar v_k$};

  \draw[decorate,decoration={brace,amplitude=4pt},line width=0.5pt] (0.0,1.2) -- (2.0,1.2)
    node[midway,above=5pt,font=\footnotesize,align=center] {future sites\\[-1pt]averaged out};
  \draw[decorate,decoration={brace,amplitude=4pt},line width=0.5pt] (2.65,1.65) -- (3.75,1.65)
    node[midway,above=5pt,font=\footnotesize] {site $k$};
  \draw[decorate,decoration={brace,amplitude=4pt},line width=0.5pt] (4.55,1.2) -- (5.85,1.2)
    node[midway,above=5pt,font=\footnotesize,align=center] {past sites\\[-1pt]already drawn};

  \node[anchor=west,font=\footnotesize] at (6.6,0.35)
    {$q(v_k\mid v_1,\dots,v_{k-1})=\gamma(v_k)\,\dfrac{v_k^\dagger G\,v_k}{\Tr G}$};
  \node[anchor=west,font=\footnotesize] at (6.6,-0.55)
    {$G_{ll'}=b_l^\dagger R_k\,b_{l'},\qquad b_l=M^{(k)}_l a_{k-1}$};
\end{scope}
\end{tikzpicture}
\caption{(a) A sequential source. One or several emitters, whose joint memory (Definition~\ref{def:memory}) bounds the bond dimension $\chi$ of the emitted state ($\chi=2$ for a single spin, $\chi\le2^k$ for $k$ spins, which may interact; photons stored in a delay line also count as memory), emit $n$ photons, one per site; each photon occupies one of the $L$ modes of its site, recorded by the label $x_k$ (for instance time bins, polarisations or spatial modes). The photons enter an $m$-mode interferometer $U$, which may act on all of these modes, and are detected in an output pattern $\mathbf t$. The temporal train reflects how the emitters generate the state; the estimator only requires each photon to occupy its own set of modes, which may equally be spatial, spectral or polarisation modes. Underneath, the same input written as a matrix-product state (Definition~\ref{def:mps}): one tensor $M^{(k)}$ per emission, linked by bonds of dimension $\chi$ and closed by the initial memory state $\ket{s_0}$ and by $\bra f$. After the last emission the memory is either measured and found in $\ket f$ (projected memory) or discarded (traced memory, Sec.~\ref{sec:traced}). Read from right to left, the chain follows the emission order. (b) One step of the sampler of Theorem~\ref{thm:is}. The top and bottom rows are the state and its complex conjugate. The values $v_1,\dots,v_{k-1}$ already drawn are contracted into the left vector $a_{k-1}$, all future sites are averaged into the right environment $R_k$, and $v_k$ is drawn from the conditional density~\eqref{eq:cond}. The diagram keeps the order of Eq.~\eqref{eq:mps}, in which matrices act from right to left, so the left vector, which collects the sites already drawn, appears on the right.}
\label{fig:schematic}
\end{figure}
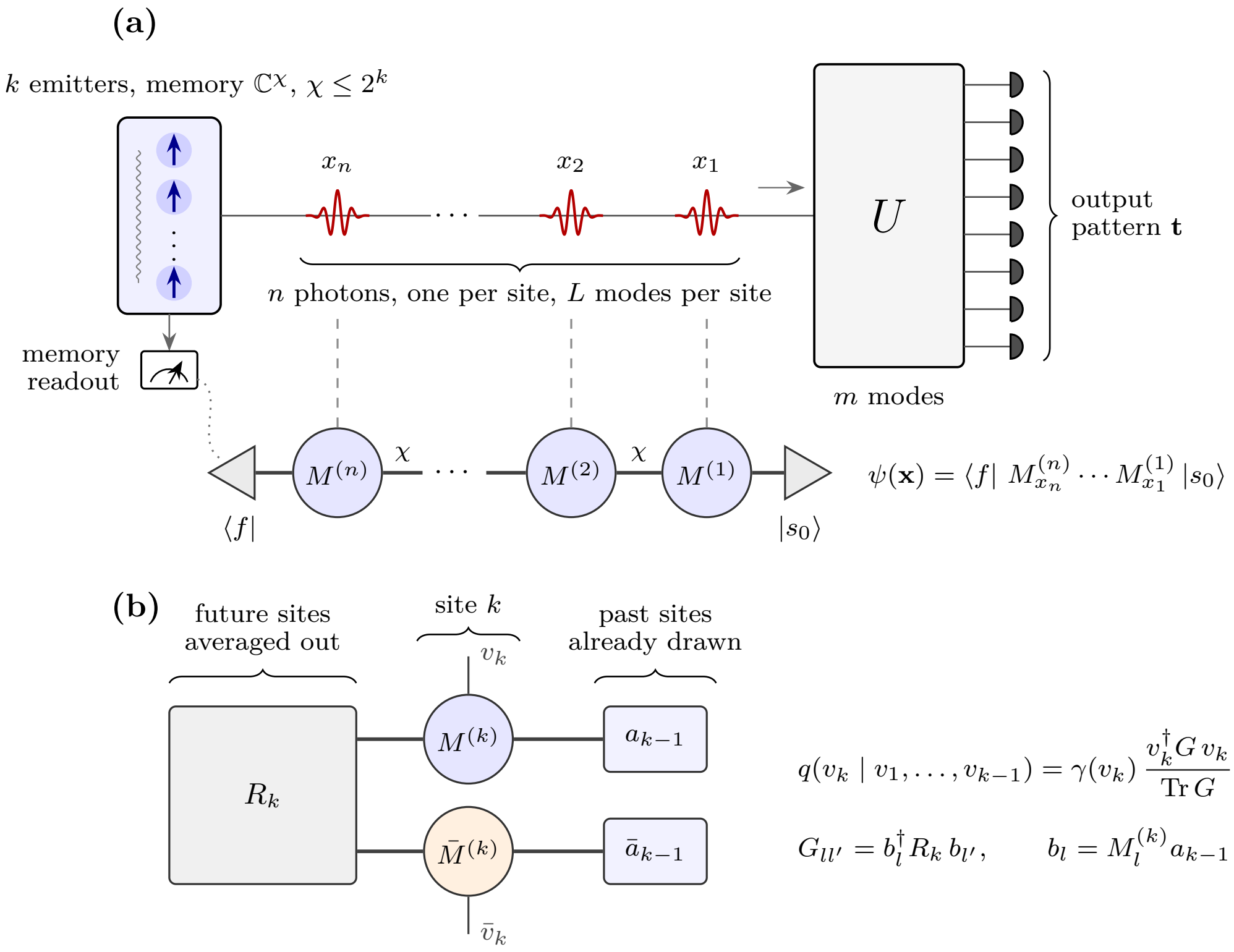

We show that it can, whenever the input is a matrix-product state (MPS) of moderate bond dimension $\chi$. Every state generated sequentially by an emitter of dimension $d$ is an MPS with $\chi\le d$, and conversely every MPS can be generated in this way \cite{Schon}. Our results are stated for any photonic MPS, whatever its physical origin, and apply in particular to all emitter-based sources. The key observation is that the importance distribution of the Lim--Oh estimator is, for an MPS input, a continuous MPS density. It can be sampled exactly one site at a time, by the perfect-sampling method of Ferris and Vidal \cite{FerrisVidal} run in a Gaussian measurement basis (Fig.~\ref{fig:schematic}). The resulting estimator keeps the attractive feature of Gurvits's method: its second moment equals the squared norm of the input, whatever the interferometer, the output pattern and the amount of entanglement. Each draw costs $O(nL\chi^2+nL^3+mn)$ operations after a one-off $O(nL\chi^3)$ preprocessing. Reaching an additive error $\varepsilon$ with probability at least $1-\delta$ takes $O(\|\psi\|^2\varepsilon^{-2}\log(1/\delta))$ draws (Sec.~\ref{sec:tails}), so the total cost is
\[
 O\Big(nL\chi^3+\big(nL\chi^2+nL^3+mn\big)\,\frac{\|\psi\|^2\log(1/\delta)}{\varepsilon^2}\Big)
\]
operations, with $O(nL\chi^2+m^2)$ memory for the input and the stored environments. The cost is polynomial in $\chi$ and in $1/\varepsilon$.

For emitter-generated light, $\chi$ is bounded by the dimension of the source's retained quantum memory (Definition~\ref{def:memory}, with examples in Table~\ref{tab:examples}): a single spin gives $\chi=2$ however long the cluster state is, and $k$ qubits of memory, whether emitters or photons held in a delay line, give $\chi\le2^k$, so the estimator stays efficient as long as $k=O(\log n)$ (Sec.~\ref{sec:consequences}). This extends the known frontier of efficient classical additive estimation from product inputs to all inputs with one photon per site and polynomial bond dimension in some ordering of the sites.

\paragraph{Contributions.}
\begin{enumerate}
\item The main result is an exact, site-by-site sampler for the importance density of an MPS input, with explicit costs in $n$, $m$, $L$, $\chi$, $\varepsilon$ and $\delta$ (Theorem~\ref{thm:is}(iii), Algorithm~\ref{alg:sampler}, Sec.~\ref{sec:tails}). Each local conditional is a mixture of at most $L$ Gaussian densities in which one coordinate has a $\mathrm{Gamma}(2,1)$ radial law. Combined with the coherent-state estimator of Lim and Oh, whose unbiasedness and state-independent second moment we restate in the fixed-photon-number setting (Proposition~\ref{thm:identity}, Theorem~\ref{thm:is}(i)--(ii)) \cite{LimOh}, it gives an efficient additive estimator for entangled inputs. Continuous Born-MPS densities and their sequential sampling are studied in general by Meiburg et al.\ \cite{Meiburg}; that work does not consider this Gaussian density, its explicit local rule, or its use in an optical estimator with a variance guarantee.
\item Extensions to traced emitters, overlaps and kernel entries, specified marginal probabilities, vacuum emissions, known loss and several photons per site (Sec.~\ref{sec:ext} and the Supplementary Material).
\item A discussion of what these additive estimates mean compared with the shot noise of an experiment, of the heavy tails of the estimator, and of how much quantum memory an advantage would need (Secs.~\ref{sec:stats} and~\ref{sec:consequences}).
\item An open-source implementation tested against brute-force enumeration, and a Lean~4 formalisation of the identities and sampling laws (Sec.~\ref{sec:numbers}).
\end{enumerate}

\paragraph{Outline.} Section~\ref{sec:background} recalls how additive estimation works for product inputs and why entanglement is an obstacle. Section~\ref{sec:model} defines photonic MPSs, explains what the bond dimension measures, and shows how emitter-generated states fit in. Sections~\ref{sec:identity} and~\ref{sec:sampler} give the identity and the sampler, and Sec.~\ref{sec:ext} the other quantities that can be estimated. Sections~\ref{sec:stats} and~\ref{sec:consequences} discuss accuracy and the consequences for quantum advantage. Section~\ref{sec:numbers} reports the numerical checks and the formalisation, and Sec.~\ref{sec:open} concludes. 

Throughout, a \emph{draw} is one evaluation of the Monte-Carlo variable, a \emph{sample} an output pattern of the boson sampler, and a \emph{shot} a run of the device.

\section{Background: additive estimation for product inputs}
\label{sec:background}

A passive linear-optical circuit on $m$ modes is described by an $m\times m$ unitary $U$. We write an $n$-photon Fock state by its occupations, $\bs=(s_1,\dots,s_m)$ with $s_j$ photons in mode $j$ and $\sum_js_j=n$, and set $\bs!=\prod_js_j!$. For a Fock input $\bs$ and output $\bt=(t_1,\dots,t_m)$,
\[
 \braket{\bt|U|\bs}=\frac{\per(U[\bt,\bs])}{\sqrt{\bs!\,\bt!}},
\]
where $U[\bt,\bs]$ is the $n\times n$ submatrix of $U$ in which row $j$ is repeated $t_j$ times and column $j$ is repeated $s_j$ times, and $\per(A)=\sum_{\sigma\in S_n}\prod_{i=1}^nA_{i\sigma(i)}$ is the permanent. Computing the permanent exactly is $\#$P-hard \cite{Valiant}, and this hardness underlies the sampling arguments of Aaronson and Arkhipov \cite{AA}.

Estimating a single amplitude to \emph{additive} error is, by contrast, easy. Gurvits \cite{Gurvits} writes, for an $n\times n$ matrix $A$ of operator norm $\|A\|$,
\[
 \per(A)=\E_x\Big[\prod_i(Ax)_i\prod_jx_j\Big],\qquad x\ \text{uniform on }\{\pm1\}^n,
\]
and averages the bracket over $O(1/\varepsilon^2)$ random sign vectors to obtain $\per(A)$ to within $\pm\varepsilon\|A\|^n$, each draw costing $O(n^2)$ operations. For collision-free outputs of a unitary, $\|A\|\le1$. Aaronson and Hance \cite{AH} extend this to repeated rows and columns, with an error that improves with the collisions by the factor $\sqrt{\prod_jt_j!/\prod_jt_j^{t_j}}\le1$.

Lim and Oh \cite{LimOh} give the estimator a physical form. For arbitrary input and output states, $\braket{\phi|U|\psi}$ is an integral over coherent states $\alpha$, weighted by the Husimi function $q(\alpha)=|\braket{\alpha|\psi}|^2/\pi^m$ of the input, and the variance of the resulting importance-sampling estimator is bounded independently of the state. The identity holds for every input. To use it, however, one must draw $\alpha$ from $q$ and evaluate the integrand, and both steps are efficient when the input is a \emph{product} state, because $q$ then factorises mode by mode. In earlier work the same authors approximate outcome probabilities with $s$-parameterised quasiprobability distributions, reaching multiplicative error for sufficiently classical circuits \cite{LimOh2023}. Importance sampling has also been used the other way round, with a boson sampler drawing from the importance distribution of a Monte-Carlo integral evaluated classically \cite{CorreaAnguita}. Tensor networks, finally, already appear in the classical simulation of linear optics, but in a different role: for lossy or noisy boson sampling, the \emph{output} state is represented as an MPS along the modes and then sampled \cite{OhLoss,OhMPS,ParkOh}. Here the \emph{input} is an MPS, the interferometer is arbitrary, and we estimate rather than sample.

For an entangled input the amplitude is a superposition of permanents, one for each component of the state (Eq.~\eqref{eq:amp} below), and $q$ no longer factorises. A direct evaluation involves exponentially many terms, each containing a permanent. This does not prove that every instance requires exponential time, and exact hardness and additive estimation are different questions. The rest of this paper shows that, for MPS inputs, the sum never needs to be expanded: one contracts the MPS once per draw.

\section{Entangled inputs as photonic matrix-product states}
\label{sec:model}

\subsection{Definition}
It is convenient to think of each photon as a qudit. Photon $k$ can occupy one of $L$ input modes, and which one it occupies is recorded by a label $x_k\in[L]=\{1,\dots,L\}$. For dual-rail qubits $L=2$; the labels may equally be time bins, frequencies or spatial modes. We call the $L$ modes available to photon $k$ its \emph{site}, and write them $\mu_{k,1},\dots,\mu_{k,L}$; different sites use different modes, so the $Ln$ modes $\mu_{k,l}$ are all distinct. In dual rail, for instance, $\mu_{k,l}=2(k-1)+l$.

A label word $\bx=(x_1,\dots,x_n)\in[L]^n$ then denotes the Fock state $\ket\bx$ with one photon at each site $k$, in mode
\[
 c_k(\bx)=\mu_{k,x_k},
\]
and we write $c(\bx)=(c_1(\bx),\dots,c_n(\bx))$ for the list of occupied modes; this list is what enters the permanents below. Because sites are disjoint, $\ket\bx$ never has two photons in the same mode, and different words give orthogonal Fock states. In this language an $n$-photon input is simply an $n$-qudit state $\sum_\bx\psi(\bx)\ket\bx$, and an MPS takes its usual form \cite{Schollwock}.

\begin{definition}[Photonic MPS]
\label{def:mps}
A photonic MPS of bond dimension $\chi$ is a nonzero state
\begin{equation}
  \ket\psi=\sum_{\bx\in[L]^n}\psi(\bx)\ket\bx,\qquad
  \psi(\bx)=\bra f\,M^{(n)}_{x_n}\cdots M^{(1)}_{x_1}\ket{s_0},
  \label{eq:mps}
\end{equation}
where each $M^{(k)}_l$ is a $\chi\times\chi$ complex matrix and $\ket{s_0},\bra f$ are boundary vectors in $\C^\chi$.
\end{definition}

Apart from the map $\bx\mapsto c(\bx)$ from labels to occupied modes, this is the standard open-boundary MPS, with physical index $x_k$ at site $k$. Every state admits this form for $\chi$ large enough; the boundary vectors $\ket{s_0}$ and $\bra f$ only close the chain.

\noindent\emph{Example.} Take $n=3$ dual-rail photons, $\mu_{k,l}=2(k-1)+l$, and $\chi=2$ with $M_1=\mathrm{diag}(1,0)$, $M_2=\mathrm{diag}(0,1)$ at every site, $s_0=(1,1)^T/\sqrt2$ and $f=(1,1)^T$. The product $M_{x_3}M_{x_2}M_{x_1}$ vanishes unless all labels agree, so $\psi(1,1,1)=\psi(2,2,2)=1/\sqrt2$, with $c(1,1,1)=(1,3,5)$ and $c(2,2,2)=(2,4,6)$. In occupation notation over the six modes this is the dual-rail GHZ state $(\ket{101010}+\ket{010101})/\sqrt2$, and each of its output amplitudes~\eqref{eq:amp} is a sum of two permanents. Taking instead $\chi=1$, $M_1=1$, $M_2=0$ gives the Fock input $\ket{101010}$ of standard boson sampling.
\medskip

Site-dependent bond dimensions $\chi_k\le\chi$ are allowed by padding with zeros. We do not assume that $\psi$ is normalised; the algorithm computes $\|\psi\|^2$ along the way. For an output occupation $\bt$ with $\sum_jt_j=n$, let $r=(r_1,\dots,r_n)$ list the output modes, mode $j$ appearing $t_j$ times. The amplitude is
\begin{equation}
  A_\bt=\braket{\bt|U|\psi}=(\bt!)^{-1/2}\sum_\bx\psi(\bx)\,\per\big(U[r,c(\bx)]\big),
  \label{eq:amp}
\end{equation}
where $U[r,c(\bx)]$ has rows $r$ and columns $c(\bx)$. The sum has $L^n$ terms.

We count dense arithmetic operations and ideal random draws. Exact sampling refers to the resulting probability law in exact arithmetic; a finite-precision implementation introduces numerical error that is not bounded here. The tensors $M^{(k)}_l$ and boundary vectors are specified inputs, rather than a preparation procedure that the classical algorithm must reproduce.

\subsection{What the bond dimension measures}
\label{sec:bond}
The bond dimension $\chi$ is a property of the photonic state itself, not of the device that produced it. It is a statement about \emph{bipartite} entanglement across cuts \emph{in site order}. If $\ket\psi$ has the form~\eqref{eq:mps}, its Schmidt rank across the cut between sites $1,\dots,k$ and $k+1,\dots,n$ is at most $\chi$; conversely, the smallest $\chi$ for which~\eqref{eq:mps} holds is the largest of these $n-1$ Schmidt ranks \cite{Vidal}. It says nothing about cuts that interleave the sites.

The distinction matters for cluster states, which are often described as maximally entangled. The linear cluster state $\prod_k\mathrm{CZ}_{k,k+1}\ket+^{\otimes n}$ (one qubit per site, $L=2$, labels written $0,1$) has Schmidt rank $2$ across every cut in chain order, since each such cut breaks a single CZ bond, and an explicit MPS with $\chi=2$ is $M_0=\tfrac1{\sqrt2}\left(\begin{smallmatrix}1&1\\0&0\end{smallmatrix}\right)$, $M_1=\tfrac1{\sqrt2}\left(\begin{smallmatrix}0&0\\1&-1\end{smallmatrix}\right)$ at every site, with $s_0=(1,0)^T$ and $f=(1,1)^T$. Across the cut separating even from odd sites, however, every bond is broken and the Schmidt rank is $2^{\lfloor n/2\rfloor}$. A GHZ state, in contrast, has Schmidt rank $2$ across every bipartition. The large multipartite entanglement of the cluster state is real, but it is invisible to the chain cuts that determine $\chi$.

The order of the sites is ours to choose. Reordering the sites only relabels input modes, and that relabelling can be absorbed into $U$. The relevant quantity for our algorithm is therefore the smallest bond dimension over all site orders, and for emitter-generated states the emission order is the natural candidate.

\subsection{States generated by quantum emitters}
\label{sec:emitters}
A quantum emitter, such as a quantum dot, an atom in a cavity or a superconducting qubit coupled to a waveguide, can be driven to emit photons one at a time while a long-lived internal degree of freedom, typically a spin, stays coherent. If the state of each emitted photon depends on that internal state, every emission entangles the photon with the emitter, which passes the entanglement on to the next photon. Sch\"on et al.\ showed that this mechanism produces precisely matrix-product states, with bond dimension set by the dimension of the emitter's memory \cite{Schon,Schon2007}. Lindner and Rudolph proposed to generate long linear cluster states in this way from a single quantum-dot spin rotated between emissions \cite{LindnerRudolph}; coupled emitters can produce two-dimensional cluster states \cite{Economou2010}, and a few emitters suffice for the tree-like states used in all-photonic repeaters \cite{Buterakos}.

These proposals have since been realised, and this is the main motivation for considering MPS inputs. Linear cluster states have been produced from the spin of a quantum dot \cite{Schwartz,Cogan}, with high rates and indistinguishable photons \cite{Coste}, and in reconfigurable form \cite{Huet}; an optical delay line can also play the role of the memory \cite{Istrati}. A single atom in a cavity has produced GHZ and linear cluster states of more than ten photons \cite{Thomas}, and two atoms in the same cavity have been used to fuse such states into larger graph states \cite{Thomas2024}. Superconducting circuits emit multipartite-entangled microwave photons \cite{Besse}, including two-dimensional cluster states generated with time-delayed feedback \cite{Ferreira}. These sources already deliver entangled multiphoton inputs to linear-optical circuits, and the question is whether such inputs change what can be computed classically.

We now give the formal description.
A sequential source carries an internal state in $\C^d$, its memory, from one emission to the next. At emission $k$ an isometry entangles the memory with a photon in the modes of site $k$; writing $M^{(k)}_l$ for the memory operator that accompanies a photon with label $l$, the joint state of photons and memory after $n$ emissions is $\ket\Psi=\sum_\bx\ket\bx\otimes M^{(n)}_{x_n}\cdots M^{(1)}_{x_1}\ket{s_0}$. This is Eq.~\eqref{eq:mps} with $\chi=d$ and an open right boundary. One spin gives $\chi=2$, and the linear cluster and caterpillar states produced by a single spin are of this form. Several emitters are described by their joint memory. Since this notion is central to what follows, we state it separately.

\begin{definition}[Memory of a sequential source]
\label{def:memory}
The \emph{memory} of a sequential source is the set of quantum degrees of freedom that are carried from one emission to the next and can influence later photons: the internal states of all emitters, and any photons held in a delay line before being fed back. If the memory has dimension $d$, for instance $d=2^k$ for $k$ qubits, the emitted photonic state is a photonic MPS (Definition~\ref{def:mps}) in emission order with $\chi\le d$.
\end{definition}

The memory is therefore a property of the source, and $\chi$ a property of the state it emits; the first bounds the second. Two spins give $\chi\le4$, $k$ two-level emitters give $\chi\le2^k$, and a delay of $\tau$ time bins \cite{Istrati,Ferreira} adds up to $\tau$ qubits. Conversely, the smallest number of emitters able to produce a given graph state in a given emission order is set by its entanglement across the cuts in that order \cite{LiEconomouBarnes}, that is, by $\log_2\chi$. The bound concerns the memory available, and a given state may use less of it. Table~\ref{tab:examples} gives examples on both sides of the threshold $k=O(\log n)$ discussed in Sec.~\ref{sec:consequences}.

\begin{table}[htbp]
\centering
\caption{Memory and bond dimension of some $n$-photon states, in the emission order indicated. The cost is that of the estimator to fixed $\varepsilon$ and $\delta$, with $m$ and $L$ polynomial in $n$.}
\label{tab:examples}
\begin{tabular}{lccc}
\toprule
State (emission order) & memory $k$ (qubits) & $\chi$ & cost\\
\midrule
Fock or product input & $0$ & $1$ & $\operatorname{poly}(n)$\\
GHZ, linear cluster (one spin) & $1$ & $2$ & $\operatorname{poly}(n)$\\
2D cluster strip of width $w$ (by columns) & $\Theta(w)$ & $2^{\Theta(w)}$ & $\operatorname{poly}(n)\,2^{O(w)}$\\
\quad with $w=O(\log n)$ & $O(\log n)$ & $\operatorname{poly}(n)$ & $\operatorname{poly}(n)$\\
\quad with $w=\sqrt n$ (square lattice) & $\Theta(\sqrt n)$ & $2^{\Theta(\sqrt n)}$ & $2^{O(\sqrt n)}$\\
Generic $n$-qubit state & $\Theta(n)$ & $2^{\Theta(n)}$ & $2^{O(n)}$\\
\bottomrule
\end{tabular}
\end{table}

At the end of the sequence the memory can be treated in two ways.

\noindent\emph{Projected memory.} If the memory is measured in a basis containing a normalised state $\ket f$ and found in that state, the photonic state is the pure, in general unnormalised, state~\eqref{eq:mps}. Then $|A_\bt|^2$ is the \emph{joint} weight of the projection outcome and the output, and the probability conditional on the projection is $|A_\bt|^2/\|\psi\|^2$. All draw counts below are stated for $A_\bt$ and $|A_\bt|^2$; the conditional version divides by $\|\psi\|^2$, which the algorithm computes.

\noindent\emph{Traced memory.} If the memory is discarded, the photonic state is $\rho=\sum_e\ket{\psi_e}\!\bra{\psi_e}$, where $\ket{\psi_e}$ is the state~\eqref{eq:mps} with $\bra f=\bra e$ and $e$ runs over an orthonormal basis of $\C^\chi$. The output probability is $p(\bt)=\sum_e|A_{\bt,e}|^2$, with $A_{\bt,e}$ the amplitude~\eqref{eq:amp} of $\psi_e$. It is a probability when the sequence of emissions is normalised: $\Tr\rho_{\mathrm{mem}}=1$ with $\rho_{\mathrm{mem}}=\sum_\bx M_\bx\ket{s_0}\!\bra{s_0}M_\bx^\dagger$ and $M_\bx=M^{(n)}_{x_n}\cdots M^{(1)}_{x_1}$. This holds whenever $\|s_0\|=1$ and each emission is an isometry from the memory to the memory and one photon, $\sum_lM^{(k)\dagger}_lM^{(k)}_l=\id$. Section~\ref{sec:traced} treats this case.

\subsection{Vacuum emissions and loss}
\label{sec:vacuum}
The model extends to emissions that may be empty (unoccupied time bins, or an emission probability below one). Each site then gains a vacuum operator $M^{(k)}_0$, the input spans several photon numbers, and the sampler of Sec.~\ref{sec:sampler} needs one rejection step per site, accepted with probability exactly $1/2$ (Supplementary Material, Sec.~\ref{app:vacuum}). Sites holding several photons can be treated in the same way when their local Fock support is small (Sec.~\ref{app:local-fock}). A known linear loss, uniform or mode-dependent, can be folded into a sub-unitary $U$ when we estimate joint weights for detecting all $n$ input photons. It leaves the identity unchanged and can only lower the second moment (Sec.~\ref{app:loss}). Unresolved loss, with a trace over lost photons and several photon-number sectors, is not covered.

\section{The Gaussian identity}
\label{sec:identity}

The starting point is to write the amplitude~\eqref{eq:amp} as an average over a Gaussian random vector, in a form that does not refer to the individual permanents. Let $\gamma(x)=\pi^{-m}e^{-\|x\|^2}$ be the standard complex Gaussian density on $\C^m$: the entries $x_a$, $a\in[m]$, are independent and circularly symmetric, with $\E|x_a|^2=1$, so that $\E[\bar x_a^{\,p}x_a^{\,q}]=p!\,\delta_{pq}$. We use the same symbol for the standard Gaussian density on the space of a single site ($\C^L$) or of the memory ($\C^\chi$). Define
\begin{equation}
  \Phi(x)=\braket{X|\psi}=\sum_\bx\psi(\bx)\prod_{k=1}^n\overline{x_{c_k(\bx)}},\qquad
  \ket X=\bigotimes_{k=1}^n\;\sum_{l=1}^{L}x_{\mu_{k,l}}\ket l,\qquad
  g_\bt(x)=\prod_{i=1}^n(Ux)_{r_i}.
  \label{eq:PhiG}
\end{equation}
Here $\Phi(x)$ is the overlap of the input with the product vector $\ket X$; for an MPS it is obtained by contracting the tensors with $\ket X$ one site at a time, in $O(nL\chi^2)$ operations. The output factor $g_\bt(x)$ only needs the $n$ entries $(Ux)_{r_i}$, which cost $O(mn)$.

\begin{proposition}[Gaussian identity]
\label{thm:identity}
For every $m\times m$ matrix $U$, every $n$-photon state with at most one photon per mode (in particular~\eqref{eq:mps}; in general, $\bx$ indexes its Fock components and $c(\bx)$ lists their occupied modes) and every output $\bt$,
\begin{equation}
  \sum_\bx\psi(\bx)\,\per\big(U[r,c(\bx)]\big)=\E_{x\sim\gamma}\big[\Phi(x)\,g_\bt(x)\big],\qquad\text{i.e.}\qquad A_\bt=(\bt!)^{-1/2}\,\E_\gamma[\Phi\,g_\bt].
  \label{eq:identity}
\end{equation}
The same holds with $x$ uniform on $\{\pm1\}^m$ (Rademacher), with no conjugation in $\Phi$.
\end{proposition}

\begin{proof}
\emph{The two forms of~\eqref{eq:identity}.} By linearity, $A_\bt=\braket{\bt|U|\psi}=\sum_\bx\psi(\bx)\braket{\bt|U|\bx}$. Each $\ket\bx$ is a Fock state with at most one photon per mode, so the formula of Sec.~\ref{sec:background} gives $\braket{\bt|U|\bx}=(\bt!)^{-1/2}\per\big(U[r,c(\bx)]\big)$: here $\bs!=1$, and $U[\bt,\bs]$ is $U[r,c(\bx)]$ up to the order of its rows and columns, which does not change the permanent. This is~\eqref{eq:amp}, and it shows that the two forms of~\eqref{eq:identity} are equivalent. It remains to prove the left-hand one.

\emph{Idea.} Each conjugated variable $\bar x_c$ in $\Phi$ stands for an input photon in mode $c$, each factor $(Ux)_{r}=\sum_dU_{rd}x_d$ in $g_\bt$ stands for an output photon in mode $r$ that may come from any input mode $d$, and the Gaussian average pairs every output photon with exactly one input photon.

\emph{Two photons.} Take one input photon in each of the modes $a\ne b$, so that $\Phi=\bar x_a\bar x_b$, and two output rows $r,s$. Expanding,
\[
 \E\big[\bar x_a\bar x_b\,(Ux)_r(Ux)_s\big]=\sum_{d,d'}U_{rd}U_{sd'}\,\E\big[\bar x_a\bar x_b\,x_dx_{d'}\big].
\]
Since the modes are independent and $\E[\bar x_a^{\,p}x_a^{\,q}]=p!\,\delta_{pq}$, the expectation on the right is $1$ if $\{d,d'\}=\{a,b\}$ and $0$ otherwise: a mode that is not occupied must not appear, and each occupied mode must appear exactly once. Two terms survive, $(d,d')=(a,b)$ and $(b,a)$, giving
\[
 \E\big[\bar x_a\bar x_b\,(Ux)_r(Ux)_s\big]=U_{ra}U_{sb}+U_{rb}U_{sa}=\per\begin{pmatrix}U_{ra}&U_{rb}\\U_{sa}&U_{sb}\end{pmatrix},
\]
the two indistinguishable ways of sending the input photons to the output rows.

\emph{General case.} Since $\Phi(x)=\sum_\bx\psi(\bx)\prod_k\bar x_{c_k(\bx)}$ is linear in $\psi$, it is enough to treat one word with occupied modes $c=(c_1,\dots,c_n)$, which are distinct by assumption. Expanding $g_\bt(x)=\prod_{i=1}^n\sum_dU_{r_id}x_d$ gives one term for each choice $d=(d_1,\dots,d_n)$ of an input mode for every output row, with coefficient $\prod_iU_{r_id_i}$. Exactly as for two photons, $\E\big[\prod_k\bar x_{c_k}\prod_ix_{d_i}\big]$ equals $1$ if every mode of $c$ is chosen exactly once and no other mode is chosen, i.e.\ if $d_i=c_{\sigma(i)}$ for a permutation $\sigma$, and $0$ otherwise. This is where distinct occupied modes are used: each conjugated variable appears at most once, so each surviving term has weight $1$. Hence
\[
 \E_\gamma\Big[\prod_k\bar x_{c_k}\,g_\bt(x)\Big]=\sum_{\sigma\in S_n}\prod_{i=1}^nU_{r_ic_{\sigma(i)}}=\per\big(U[r,c]\big).
\]
Repeated output rows cause no difficulty, since the sum runs over choices for each row. Summing over words with weights $\psi(\bx)$ gives the left-hand form of~\eqref{eq:identity}.

\emph{Signs.} For the last statement of the proposition, take $x$ uniform on $\{\pm1\}^m$; the variables are real, so the conjugation in $\Phi$ plays no role. Now $\E[x_a^k]$ is $1$ for even $k$ and $0$ otherwise, so each mode must appear an even number of times in $\prod_kx_{c_k}\prod_ix_{d_i}$. Every mode of $c$ therefore has to be chosen at least once; with only $n$ choices for $n$ modes, each is chosen exactly once, and the same permutations survive.
\end{proof}

Proposition~\ref{thm:identity} is the fixed-photon-number form of the coherent-state identity of \cite{LimOh}: $\ket X$ is the projection of the unnormalised coherent-state generating vector with amplitude $x$ onto the sector with one photon in each site, and $\gamma$ is the Gaussian factor of the Husimi measure. It uses no MPS structure. Averaging $\Phi g_\bt$ directly under $\gamma$ is unbiased, but its variance involves $\E_\gamma|\Phi|^2|g_\bt|^2$, which is not controlled. As in \cite{Gurvits,LimOh}, we move the input factor into the sampling measure; for an MPS the resulting measure can be sampled exactly, as shown next.

\section{The sequential importance sampler}
\label{sec:sampler}

Physically, sampling $x$ from $\gamma|\Phi|^2$ amounts to measuring every site of the input in an (overcomplete) Gaussian product basis and recording the outcomes. For an MPS such a measurement can be simulated site by site: one keeps a $\chi$-dimensional vector summarising the outcomes already drawn, and a $\chi\times\chi$ matrix summarising all possible futures. This is the perfect-sampling algorithm of Ferris and Vidal \cite{FerrisVidal}, with the computational basis replaced by Gaussian product vectors.

\begin{theorem}[Importance-sampled estimator]
\label{thm:is}
Let $q(x)=\gamma(x)|\Phi(x)|^2/\|\psi\|^2$. Then
\begin{enumerate}
\item[(i)] $q$ is a probability density on $\C^m$;
\item[(ii)] under $x\sim q$, the variable $Z(x)=\|\psi\|^2\,g_\bt(x)/\overline{\Phi(x)}$ (defined off the $q$-null set $\{\Phi=0\}$) satisfies, for every unitary $U$, every $\bt$ and every $\psi$,
  \[
    \E_q[Z]=\sqrt{\bt!}\,A_\bt,\qquad
    \E_q|Z|^2=\|\psi\|^2\,\bt!\,,
  \]
  so that $\hat Z=Z/\sqrt{\bt!}$ is unbiased for $A_\bt$ with $\E|\hat Z|^2=\|\psi\|^2$, independent of $U$, $\bt$ and the detailed input state;
\item[(iii)] for the MPS~\eqref{eq:mps}, $q$ can be sampled exactly, and $\Phi(x)$ and $\|\psi\|^2$ evaluated, in $O(nL\chi^2+nL^3)$ operations per draw after $O(nL\chi^3)$ preprocessing per state; $g_\bt$ adds $O(mn)$ per draw.
\end{enumerate}
Consequently, for $N$ independent draws $x_1,\dots,x_N\sim q$, $\hat A=N^{-1}\sum_{j=1}^N\hat Z(x_j)$ satisfies $\E|\hat A-A_\bt|^2\le\|\psi\|^2/N$, and for two independent such estimates $\hat A_1\overline{\hat A_2}$ is unbiased for $|A_\bt|^2$ with mean-square error $\le\|\psi\|^4/N$.
\end{theorem}

Parts (i) and (ii) hold for any input with at most one photon per mode; they are the fixed-photon-number form of the estimator of Lim and Oh and of its state-independent variance bound \cite{LimOh}. Part (iii) uses the MPS form and is the new ingredient.

Why not reuse the product-state estimator word by word? A natural alternative draws a component $\bx$ with probability $|\psi(\bx)|^2/\|\psi\|^2$, which perfect sampling of the MPS in the computational basis does in $O(nL\chi^2)$ operations, and then estimates the single amplitude $\braket{\bt|U|\bx}$ with Gurvits's method. Reweighting by $\|\psi\|^2/\overline{\psi(\bx)}$ makes this unbiased as well, but each word now contributes its own variance, and the second moment becomes $\|\psi\|^2$ times the number of words in the support of $\psi$. This is harmless for a GHZ state, which has two words, but grows as $2^n$ for a linear cluster state, whose amplitudes are all nonzero. The same holds for any approach that writes $\ket\psi$ as a sum of product states and applies Gurvits's method to each: the cost grows with the number of terms, which for a linear cluster state is at least its Schmidt rank across the even--odd cut, $2^{\lfloor n/2\rfloor}$ (Sec.~\ref{sec:bond}), although $\chi=2$. The Gaussian change of measure of Theorem~\ref{thm:is} avoids this because it treats all words at once: their cross terms cancel in $\E_\gamma|\Phi|^2$, and the second moment does not depend on how many words the state contains.

\begin{proof}
Write $r_\psi=\|\psi\|^2$. We first explain the change of measure, then how to sample it.

\paragraph{Normalising the input weight.}
Different input words use different sets of modes. Their Gaussian monomials are orthogonal, so all cross terms disappear in $\E_\gamma|\Phi|^2$:
\[
 \E_\gamma|\Phi(x)|^2=\sum_\bx|\psi(\bx)|^2=r_\psi.
\]
This proves (i). Physically, integrating the product-vector measurement sums over all its outcomes and returns the norm of the input state.

\paragraph{Cancelling the input from the second moment.}
Multiplying the estimator by its sampling density gives
\[
 q(x)Z(x)=\gamma(x)\Phi(x)g_\bt(x),\qquad
 q(x)|Z(x)|^2=r_\psi\gamma(x)|g_\bt(x)|^2
\]
wherever $\Phi(x)\ne0$. Integrating the first equality and using Proposition~\ref{thm:identity} gives $\E_q[Z]=\E_\gamma[\Phi g_\bt]=\sqrt{\bt!}\,A_\bt$, the unbiasedness claim of (ii). For the second equality, the set $\{\Phi=0\}$ is the zero set of a nonzero polynomial, so it has Lebesgue, hence Gaussian, measure zero, and integrating gives $\E_q|Z|^2=r_\psi\,\E_\gamma|g_\bt|^2$. For the sign variant this step fails, since a polynomial can vanish on sign vectors with positive probability (Sec.~\ref{sec:rademacher}).

It remains to compute $\E_\gamma|g_\bt|^2$, and here we use that $U$ is unitary, as it is for a lossless linear-optical circuit. Since $\gamma$ depends on $x$ only through $\|x\|=\|Ux\|$, the vector $y=Ux$ has the same distribution as $x$, and $g_\bt(x)=\prod_jy_j^{t_j}$. By independence and the moment rule,
\[
 \E_\gamma|g_\bt(x)|^2
 =\prod_j\E|y_j|^{2t_j}=\prod_jt_j!=\bt!.
\]
Dividing by $\bt!$ gives $\E|\hat Z|^2=r_\psi$, which completes the proof of (ii): the input enters the second moment only through its norm. For a lossy, sub-unitary $U$ the same quantity is at most $\bt!$ (Sec.~\ref{app:loss}).

\paragraph{Sampling one site at a time.}
It remains to draw $x$ from $q$. The function $\Phi$ depends only on the modes used by the sites, so $q$ is the product of $\gamma$ on the unused modes, which are drawn as independent standard Gaussians, and a density on the site variables. For site $k$ write $v_k=(v_{k,1},\dots,v_{k,L})\in\C^L$ with $v_{k,l}=\bar x_{\mu_{k,l}}$; conjugation preserves $\gamma$, so under $\gamma$ the $v_k$ are independent standard Gaussian vectors. By the chain rule, it suffices to draw $v_1$ from its marginal under $q$, then $v_2$ from its law conditional on $v_1$, and so on. We therefore need the marginal density of the first $k$ site vectors. Two objects make it computable.

\emph{The left vector} $a_k\in\C^\chi$ is the memory vector after the first $k$ sites, given the values $v_1,\dots,v_k$ already drawn:
\[
 a_0=s_0,\qquad a_k=\sum_{l=1}^Lv_{k,l}\,M^{(k)}_l\,a_{k-1},\qquad \Phi=f^\dagger a_n .
\]
This is~\eqref{eq:PhiG} evaluated by multiplying the matrices from the right, one site at a time.

\emph{The right environment} $R_k$ is a $\chi\times\chi$ positive semidefinite matrix that summarises all possible futures. It is defined backwards by
\[
 R_n=ff^\dagger,\qquad R_{k-1}=\sum_{l=1}^LM^{(k)\dagger}_l\,R_k\,M^{(k)}_l ,
\]
and does not depend on $x$, so it is computed once per state.

\paragraph{Averaging over the future.}
The key property is that averaging $|\Phi|^2$ over the sites that have not yet been drawn leaves a quantity that depends on the past only through $a_k$:
\begin{equation}
 \int\Big(\prod_{k'>k}\gamma(v_{k'})\,dv_{k'}\Big)|\Phi|^2=a_k^\dagger R_k\,a_k .
 \label{eq:future}
\end{equation}
For $k=n$ there is nothing to average, and $|\Phi|^2=a_n^\dagger ff^\dagger a_n=a_n^\dagger R_na_n$. Suppose~\eqref{eq:future} holds at $k$, and average once more over $v_k$. Writing $b_l=M^{(k)}_la_{k-1}$, so that $a_k=\sum_lv_{k,l}b_l$,
\[
 \int\gamma(v_k)\,a_k^\dagger R_ka_k\,dv_k=\sum_{l,l'}\E\big[\bar v_{k,l}v_{k,l'}\big]\,b_l^\dagger R_kb_{l'}=\sum_lb_l^\dagger R_kb_l=a_{k-1}^\dagger R_{k-1}a_{k-1},
\]
because $\E[\bar v_{k,l}v_{k,l'}]=\delta_{ll'}$ removes the cross terms. This is~\eqref{eq:future} at $k-1$. At $k=0$ it gives $\|\psi\|^2=\E_\gamma|\Phi|^2=s_0^\dagger R_0s_0$, without enumerating the $L^n$ words.

\paragraph{Marginal and conditional densities.}
Dividing~\eqref{eq:future} by $\|\psi\|^2$, the marginal density of the first $k$ site vectors under $q$ is
\[
 q(v_1,\dots,v_k)=\Big(\prod_{k'\le k}\gamma(v_{k'})\Big)\frac{a_k^\dagger R_ka_k}{\|\psi\|^2}.
\]
The conditional density of $v_k$ given the past is the ratio of two consecutive marginals. The Gaussian factors of the past and $\|\psi\|^2$ cancel, leaving
\[
 q(v_k\mid v_1,\ldots,v_{k-1})=\gamma(v_k)\,\frac{a_k^\dagger R_ka_k}{a_{k-1}^\dagger R_{k-1}a_{k-1}} .
\]
With $b_l=M^{(k)}_la_{k-1}$ as above, the numerator is the quadratic form $v_k^\dagger Gv_k$ of the $L\times L$ matrix $G_{ll'}=b_l^\dagger R_kb_{l'}$, and the denominator is $\Tr G$:
\begin{equation}
 q(v_k\mid v_1,\ldots,v_{k-1})
 =\gamma(v_k)\,\frac{v_k^\dagger Gv_k}{\Tr G}.
 \label{eq:cond}
\end{equation}
$G$ is a Gram matrix, hence positive semidefinite. Its trace vanishes only on prefixes that have probability zero, so the conditional is always well defined when it is needed.

\paragraph{Drawing from the conditional.}
Diagonalise $G=\sum_{\alpha=1}^L\lambda_\alpha e_\alpha e_\alpha^\dagger$ with $\lambda_\alpha\ge0$ and orthonormal $e_\alpha$, and use the coordinates $c_\alpha=e_\alpha^\dagger v_k$. Since $\gamma$ is invariant under unitaries, $c$ is again standard Gaussian under $\gamma$, and~\eqref{eq:cond} becomes
\[
 \gamma(c)\sum_{\alpha=1}^L\frac{\lambda_\alpha}{\Tr G}\,|c_\alpha|^2 .
\]
This is a mixture of $L$ simple densities with weights $\lambda_\alpha/\Tr G$. In component $\alpha$, the coordinate $c_\alpha$ has density $\gamma(c_\alpha)|c_\alpha|^2$ and the other coordinates are independent standard Gaussians. In polar coordinates $c_\alpha=\sqrt s\,e^{i\theta}$, the density $\gamma(c_\alpha)|c_\alpha|^2$ becomes $s\,e^{-s}\,ds\,d\theta/(2\pi)$: $s$ follows a $\mathrm{Gamma}(2,1)$ law and $\theta$ is uniform. One draw therefore consists of
\begin{enumerate}
\item choosing $\alpha$ with probability $\lambda_\alpha/\Tr G$;
\item drawing $|c_\alpha|^2\sim\mathrm{Gamma}(2,1)$ with a uniform phase, and every other $c_{\alpha'}$ as a standard complex Gaussian;
\item setting $v_k=\sum_\alpha c_\alpha e_\alpha$, then $a_k=\sum_lv_{k,l}b_l$, and $x_{\mu_{k,l}}=\bar v_{k,l}$.
\end{enumerate}
Repeating this for $k=1,\dots,n$ draws the site variables from $q$ exactly, by the chain rule, and returns $\Phi=f^\dagger a_n$ at no extra cost.

\paragraph{Cost.}
The backward recursion for $R_{n-1},\dots,R_0$ multiplies $L$ pairs of $\chi\times\chi$ matrices per site, $O(nL\chi^3)$ in total. At each site of a draw, forming the $b_l$ and $R_kb_l$ costs $O(L\chi^2)$, the matrix $G$ costs $O(L^2\chi)$, and its diagonalisation $O(L^3)$; since $L^2\chi\le(L\chi^2+L^3)/2$, this is $O(L\chi^2+L^3)$ per site. This proves (iii).

\paragraph{From one draw to an error bound.}
Independence gives $\E|\hat A-A_\bt|^2=(r_\psi-|A_\bt|^2)/N\le r_\psi/N$. To estimate a probability without squaring a noisy estimate, use two independent averages. Put $a=|A_\bt|^2$ and $v=r_\psi-a$. Their product has mean $a$ and mean-square error
\[
  \frac{2av}{N}+\frac{v^2}{N^2}
  \le\frac{2av+v^2}{N}
  =\frac{r_\psi^2-a^2}{N}\le\frac{r_\psi^2}{N}.
\]
The real part is also unbiased and has no larger mean-square error.
\end{proof}

\begin{algorithm}[One Gaussian importance draw]
\label{alg:sampler}
Preprocess once: $R_n=ff^\dagger$, $R_{k-1}=\sum_lM_l^{(k)\dagger}R_kM_l^{(k)}$, and $r_\psi=s_0^\dagger R_0s_0$.
\begin{enumerate}
\item Set $a=s_0$. Draw independent standard complex Gaussian entries $x_a$ on modes outside the sites. (Setting them to zero instead also gives an unbiased estimator, with a smaller second moment when $m\gg Ln$; Remark~\ref{rem:zero}.)
\item For each site $k=1,\ldots,n$, perform steps 3--5.
\item Form $b_l=M_l^{(k)}a$, $\tilde b_l=R_kb_l$ and $G_{ll'}=b_l^\dagger\tilde b_{l'}$.
\item Diagonalise $G=\sum_\alpha\lambda_\alpha e_\alpha e_\alpha^\dagger$. Choose $\alpha$ with probability $\lambda_\alpha/\Tr G$. Draw $|c_\alpha|^2\sim\mathrm{Gamma}(2,1)$ with uniform phase, and independent $c_{\alpha'}\sim\mathcal{CN}(0,1)$ for $\alpha'\ne\alpha$. Set $v=\sum_\alpha c_\alpha e_\alpha$.
\item Update $a=\sum_l v_lb_l$ and set $x_{\mu_{k,l}}=\bar v_l$.
\item After the last site, set $\Phi=f^\dagger a$ and return
\[
 \hat Z=\frac{r_\psi}{\sqrt{\bt!}\,\bar\Phi}\prod_{i=1}^n(Ux)_{r_i}.
\]
\end{enumerate}
With Gaussian unused modes and unitary $U$, $\E\hat Z=A_\bt$ and $\E|\hat Z|^2=r_\psi$.
\end{algorithm}

\begin{remark}[Sampling without diagonalisation]
\label{rem:factor}
Step~4 can avoid the eigenproblem. Factor each environment once, $R_k=C_kC_k^\dagger$ with $C_k$ a $\chi\times\kappa_k$ matrix and $\kappa_k=\operatorname{rank}R_k\le\chi$ (for instance by a pivoted Cholesky or an eigenvalue decomposition), which adds $O(n\chi^3)$ to the preprocessing. With $(w_\beta)_l=(C_k^\dagger b_l)_\beta$ for $\beta=1,\dots,\kappa_k$, the quadratic form in~\eqref{eq:cond} becomes $v^\dagger Gv=\sum_\beta|w_\beta^Tv|^2$, so the conditional density is a mixture of at most $\kappa_k$ densities $\gamma(v)|u_\beta^\dagger v|^2$, with $u_\beta=\bar w_\beta/\|w_\beta\|$ and weights $\|w_\beta\|^2/\Tr G$ (terms with $w_\beta=0$ drop out). Each of them is drawn as in step~4 with $u_\beta$ in place of the eigenvector: choose $\beta$, draw $v'$ from $\gamma$ and $c$ with $|c|^2\sim\mathrm{Gamma}(2,1)$ and uniform phase, and set $v=v'+(c-u_\beta^\dagger v')\,u_\beta$. A site then costs $O(L\chi^2)$, the $nL^3$ term disappears from Theorem~\ref{thm:is}(iii), and no eigenproblem is solved during a draw; the only factorisations are those of the $R_k$, done once per state. In our tests (Supplementary Material, Sec.~\ref{app:numbers}) the two samplers agree on all tested one- and two-site moments of $q$ and give unbiased estimates. For $L=2$ they run at comparable speed when $\chi\ge16$; the factorised sampler is faster for small $\chi$ and for encodings with many modes per photon.
\end{remark}

\section{Other quantities we can estimate}
\label{sec:ext}
An amplitude for one output pattern is rarely the quantity one actually needs. An experiment that keeps the memory unmeasured produces a mixed photonic state, applications such as kernel methods compare two states, and training a model usually involves marginal probabilities over a few output modes rather than full patterns. All three can be estimated with the same sampler, at polynomial cost. We first treat a memory that is traced out rather than projected, then replace the Fock output by a second MPS to obtain overlaps, and finally combine these overlaps to extract marginal probabilities. In each case the estimator is reused as it is, and nothing requires summing over the exponentially many output patterns.

\subsection{Traced memory}
\label{sec:traced}
Tracing out the memory means adding the probabilities associated with an orthonormal basis of final memory states. Rather than estimating all $\chi$ terms, we draw one basis state at random, with probability proportional to its weight, and reweight the pure-state estimator accordingly.

\begin{corollary}[Traced memory]
\label{cor:traced}
Let the memory be traced out, let $\{\ket e\}$ be an orthonormal basis of $\C^\chi$, and let $\psi_e$ be the state~\meqref{eq:mps} with $\bra f=\bra e$, of squared norm $r_e=\|\psi_e\|^2=\bra e\rho_{\mathrm{mem}}\ket e$, so that $\sum_er_e=\Tr\rho_{\mathrm{mem}}$. Draw $e$ with probability $r_e/\Tr\rho_{\mathrm{mem}}$, then two independent draws $\hat Z_1,\hat Z_2$ of the estimator of Theorem~\mref{thm:is} for $\psi_e$, and set
\[
 \hat P=\frac{\Tr\rho_{\mathrm{mem}}}{r_e}\,\hat Z_1\overline{\hat Z_2}.
\]
Then $\E\hat P=p(\bt)=\sum_e|A_{\bt,e}|^2$ and, with Gaussian unused modes and unitary $U$, $\E|\hat P|^2=(\Tr\rho_{\mathrm{mem}})^2$, equal to $1$ for a normalised sequence. Hence $p(\bt)$ is estimated to $\pm\varepsilon$ with $O((\Tr\rho_{\mathrm{mem}})^2/\varepsilon^2)$ draws, each costing $O(nL\chi^2+nL^3+mn)$ after at most $O(nL\chi^4)$ preprocessing (one set of right environments for each basis state that is drawn).
\end{corollary}
\begin{proof}
Basis states with $r_e=0$ are never drawn and contribute nothing to $p(\bt)$, since then $\psi_e=0$. Given $e$, the draws $\hat Z_1,\hat Z_2$ are independent, so $\E[\hat Z_1\overline{\hat Z_2}\mid e]=|A_{\bt,e}|^2$ and $\E[|\hat Z_1\hat Z_2|^2\mid e]=r_e^2$ by Theorem~\mref{thm:is}(i)--(ii). Averaging over $e$,
\[
 \E\hat P=\sum_e\frac{r_e}{\Tr\rho_{\mathrm{mem}}}\,\frac{\Tr\rho_{\mathrm{mem}}}{r_e}\,|A_{\bt,e}|^2=p(\bt),\qquad
 \E|\hat P|^2=\sum_e\frac{r_e}{\Tr\rho_{\mathrm{mem}}}\,\frac{(\Tr\rho_{\mathrm{mem}})^2}{r_e^2}\,r_e^2=(\Tr\rho_{\mathrm{mem}})^2 .
\]
The weights $r_e$ are the diagonal of $\rho_{\mathrm{mem}}$, obtained by one forward contraction $\rho_k=\sum_lM^{(k)}_l\rho_{k-1}M^{(k)\dagger}_l$ from $\rho_0=s_0s_0^\dagger$, in $O(nL\chi^3)$. The environments of Theorem~\mref{thm:is} with $f=e$ cost $O(nL\chi^3)$ for each $e$ and can be built when $e$ is first drawn; each draw then runs the sampler of Theorem~\mref{thm:is} twice with stored environments. Keeping all $\chi$ sets requires $O(n\chi^3)$ memory.
\end{proof}

A Gaussian final vector gives an alternative that needs no basis. If $z\in\C^\chi$ is standard complex Gaussian and $\psi_z$ is the state with $\bra f=\bra z$, then $\int\gamma(z)\ket z\!\bra z\,dz=\id$ gives $p(\bt)=\E_z|A_{\bt,z}|^2$, and $W=\hat Z_1\overline{\hat Z_2}$, with two independent draws for the same $z$, is unbiased. Its second moment is $\E_z\|\psi_z\|^4=(\Tr\rho_{\mathrm{mem}})^2+\Tr\rho_{\mathrm{mem}}^2\le2(\Tr\rho_{\mathrm{mem}})^2$, by Wick's theorem applied to $\|\psi_z\|^2=\bra z\rho_{\mathrm{mem}}\ket z$, and the environments must be rebuilt for every $z$, at $O(nL\chi^3)$ per draw. With environments built on demand, Corollary~\ref{cor:traced} is therefore never more expensive in operation counts, and its second moment is smaller. The reference package implements both routes, with Corollary~\ref{cor:traced} as the default; Table~\ref{tab:valid} checks the Gaussian one.

\subsection{Overlaps}
\label{sec:overlap}
An overlap seems to require summing amplitudes over every output pattern. We can instead contract the second state's MPS with the transformed Gaussian vector $Ux$, and Gaussian orthogonality performs the output sum for us.

Let $\ket\phi$ be a second photonic MPS, of bond dimension $\chi_\phi$, over any grouping of the modes into sites (not necessarily that of $\psi$; in this subsection $\mu_{k,l}$ denotes the sites of $\phi$). Then $\braket{\phi|U|\psi}=\sum_\bt\overline{\phi(\bt)}\braket{\bt|U|\psi}$ with $\phi(\bt)=\braket{\bt|\phi}$. Define, for $y\in\C^m$ and with $r(\bt)$ the rows of $\bt$,
\[
  H_\phi(y)=\sum_{\bt}\frac{\overline{\phi(\bt)}}{\sqrt{\bt!}}\,\prod_i y_{r_i(\bt)}=\overline{\braket{Y|\phi}}\quad\text{with }\ket Y=\bigotimes_k\sum_ly_{\mu_{k,l}}\ket l.
\]
For $\phi$ of the form~\meqref{eq:mps} every $\bt$ in its support is collision-free, so $\bt!=1$, and $H_\phi(y)=\sum_\bx\overline{\phi(\bx)}\prod_ky_{\mu_{k,x_k}}$ is the contraction of the \emph{complex-conjugated} tensors of $\phi$ (i.e.\ $\overline{M^{(k)}_l}$, $\bar s_0$, $\bar f$) with the product vector built from $y$, $O(nL\chi_\phi^2)$. Write $\Phi_\psi$ and $q_\psi$ for the $\Phi$ and $q$ of $\psi$. Then $\braket{\phi|U|\psi}=\E_\gamma[\Phi_\psi(x)H_\phi(Ux)]$ by Proposition~\mref{thm:identity} summed over $\bt$, and sampling $x\sim q_\psi$ and averaging $\|\psi\|^2H_\phi(Ux)/\overline{\Phi_\psi(x)}$ gives an unbiased estimate with second moment $\|\psi\|^2\,\E_\gamma|H_\phi(Ux)|^2=\|\psi\|^2\|\phi\|^2$, by unitary invariance of $\gamma$ and part (i) of Theorem~\mref{thm:is} applied to $\phi$. The overlap is therefore estimated to $\pm\varepsilon$ with $O(\|\psi\|^2\|\phi\|^2/\varepsilon^2)$ draws, each costing $O(nL\chi^2+nL\chi_\phi^2+nL^3+m^2)$; the last term computes all of $Ux$, which $H_\phi$ may need.

This covers kernel entries. For two photonic MPSs $\psi_1,\psi_2$ and two interferometers $U,U'$, the matrix $U^\dagger U'$ is again unitary, so $\braket{\psi_1|U^\dagger U'|\psi_2}$ is an overlap of the above form, and the kernel entry $|\braket{\psi_1|U^\dagger U'|\psi_2}|^2$ is estimated without bias by the product of two independent overlap estimates, as for probabilities in Theorem~\mref{thm:is}. These are the quantities used by quantum kernel methods.

\subsection{Few-mode marginals and linear statistics}
\label{sec:marginals}
An amplitude estimator does not by itself give marginal probabilities, but the overlap estimator does, through characteristic functions.

\emph{Characteristic functions as overlaps.} Let $\hat n_j$ be the number operator of output mode $j$ and $\Omega=\sum_j\omega_j\hat n_j$ a linear statistic \cite{Seron}. For $\theta\in\mathbb R$, $e^{i\theta\Omega}$ is the Fock-space action of the single-particle matrix $D_\theta=\mathrm{diag}(e^{i\theta\omega_j})$, so $V_\theta=U^\dagger D_\theta U$ is again an $m\times m$ unitary and
\[
  C_\Omega(\theta)=\braket{\psi|U^\dagger e^{i\theta\Omega}U|\psi}=\braket{\psi|V_\theta|\psi}
\]
is an overlap of the form of Sec.~\ref{sec:overlap} with $\phi=\psi$. It is estimated to $\pm\varepsilon$ with $O(\|\psi\|^4/\varepsilon^2)$ draws, each costing $O(nL\chi^2+nL^3+m^2)$. For a normalised input, characteristic functions determine probabilities; for an unnormalised one, they determine weights.

\emph{Marginals on a few modes.} For a set $S$ of $s$ output modes, the joint occupation $\bt_S\in\{0,\dots,n\}^s$ has characteristic function $C_S(\theta)=\braket{\psi|U^\dagger e^{i\theta\cdot\hat n_S}U|\psi}$, with $\theta\cdot\hat n_S=\sum_{j\in S}\theta_j\hat n_j$. Since each occupation lies between $0$ and $n$, it is enough to know $C_S$ on the grid $\theta\in\big(\tfrac{2\pi}{n+1}\mathbb Z_{n+1}\big)^s$, and the marginal is the inverse discrete Fourier transform
\[
  p(\bt_S)=(n+1)^{-s}\sum_\theta e^{-i\theta\cdot\bt_S}\,C_S(\theta)=\E_\theta\big[e^{-i\theta\cdot\bt_S}C_S(\theta)\big],
\]
with $\theta$ uniform on the grid. The Fourier phases act as a filter: their average is one when the measured occupations equal the requested ones and zero otherwise. The grid has $(n+1)^s$ points but never needs to be enumerated. One draws a single grid point $\theta$ uniformly and one overlap estimate $W_\theta$ of $C_S(\theta)$; then $e^{-i\theta\cdot\bt_S}W_\theta$ is unbiased for $p(\bt_S)$, with second moment at most $\|\psi\|^4$. Hence $p(\bt_S)$ is estimated to $\pm\varepsilon$ with $O(\|\psi\|^4/\varepsilon^2)$ draws, whatever the number $s$ of modes.

\emph{Linear statistics.} The same trick gives $\Pr[\Omega=\omega]$ when the weights $\omega_j$ are integers. After shifting the statistic by $n\min_j\omega_j$, its values lie in $\{0,\dots,n(\max_j\omega_j-\min_j\omega_j)\}$, so a one-dimensional Fourier grid of size $Q>n(\max_j\omega_j-\min_j\omega_j)$ separates them, and again only one grid point is drawn per estimate.

In both cases the result is an additive estimate of one specified probability, not an efficient way to output the whole marginal table, and for large $s$ the probabilities themselves are typically far below $\varepsilon$.

\section{Accuracy and statistical interpretation}
\label{sec:stats}

\subsection{Additive error and shot noise}
\label{sec:additive}
How do these guarantees compare with what an experiment delivers? With $N$ shots, the device estimates $p(\bt)$ with standard deviation $\sqrt{p(1-p)/N}$. With $N$ draws, two independent amplitude averages give a probability estimate with root-mean-square error at most $\|\psi\|^2/\sqrt N$, and for a normalised sequence with the memory traced out, Corollary~\ref{cor:traced} gives at most $1/\sqrt N$ from a single average, each draw using two runs of the sampler. Both errors decrease as $N^{-1/2}$, and for probabilities of order one they agree up to a constant factor.

They differ for rare outcomes. The shot noise is proportional to $\sqrt p$ and shrinks with the probability, whereas the classical error does not. In boson sampling, rare is the rule: the probability of an individual output pattern is typically exponentially small. To see the scale, take a normalised $40$-photon input in $80$ modes and a Haar-random $U$. Whatever the input state, the mean probability of a fixed output pattern is $1/\binom{119}{40}\simeq1.3\cdot10^{-32}$, since the Haar average spreads the probability uniformly over the $\binom{119}{40}$ patterns. With $N=10^5$, the shot noise is then about $4\cdot10^{-19}$ and the classical error about $3\cdot10^{-3}$. The device is far more precise in absolute terms, yet both errors exceed the probability itself by many orders of magnitude. Neither polynomially many shots nor polynomially many draws give relative precision on such probabilities, so neither can usefully report individual pattern probabilities.

What both can estimate at polynomial cost, exactly as for Fock inputs, are quantities of order one: few-mode marginals and linear statistics, overlaps and kernel entries (Secs.~\ref{sec:overlap}--\ref{sec:marginals}), and the weight of a fixed pattern when that weight is not exponentially small. These are the quantities from which loss functions and kernels are built, and for them the classical estimator matches the device up to a constant factor in the number of samples.

\subsection{Heavy tails}
\label{sec:tails}
Theorem~\ref{thm:is} controls the second moment of $\hat Z$, and that is all the error bound needs. In practice, however, one is tempted to judge the accuracy of a run from the data, using the sample variance of the draws. For this estimator that is unreliable, for two reasons.

First, the fourth moment of $\hat Z$ is in general infinite, so the sample variance itself has infinite variance. The cause is the division by $\bar\Phi$. Near a simple zero of $\Phi$ at which $g_\bt$ does not vanish, the sampling density $q\propto\gamma|\Phi|^2$ vanishes quadratically while $|\hat Z|^4$ grows as $|\Phi|^{-4}$; the integrand of $\E_q|\hat Z|^4$ therefore behaves as $|\Phi|^{-2}$, whose integral diverges logarithmically in the two real directions transverse to the zero set.

Second, the distribution of $|\hat Z|^2$ is very skewed: its typical values lie far below its mean, which is carried by rare large draws. A finite run usually misses them and underestimates the variance, sometimes by orders of magnitude. For a projected linear cluster state with $\chi=2$, $10^5$ draws gave an empirical second moment of $2\cdot10^{-5}$ at $n=40$, against the exact value $\|\psi\|^2=1/2$ (Supplementary Material, Sec.~\ref{app:tails}).

We therefore do not rely on error estimates from the data. The guarantee comes from the theorem: by Chebyshev's inequality, $|\hat A-A_\bt|\le\sqrt{\|\psi\|^2/(N\delta)}$ with probability at least $1-\delta$. When high confidence is needed, a median-of-means estimator, applied to real and imaginary parts separately, reaches error $\varepsilon$ with probability $1-\delta$ using $O(\|\psi\|^2\varepsilon^{-2}\log(1/\delta))$ draws. The same caution applies to the probability, overlap and marginal estimators of Sec.~\ref{sec:ext}, whose guarantees also rest on second-moment bounds.

\section{How much quantum memory would an advantage need?}
\label{sec:consequences}

For a sequential source whose retained quantum memory amounts to $k$ qubits, held in emitters or in photons stored in a delay line, the bond dimension in emission order is at most $2^k$ (Sec.~\ref{sec:emitters}). The cost of our estimator is polynomial in $n$, $m$, $L$, $\chi$ and $1/\varepsilon$, so substituting $\chi\le2^k$ gives a runtime $\operatorname{poly}(n,m,L,1/\varepsilon)\,2^{O(k)}$. With $m$, $L$ and $1/\varepsilon$ polynomial in $n$, this is polynomial when $k=O(\log n)$ and subexponential when $k=o(n)$. Table~\ref{tab:examples} gives examples of each regime. In particular, the tasks considered here can only become exponentially hard if the memory grows linearly with the number of photons; for a memory that grows faster than logarithmically but sublinearly, our bound is superpolynomial but subexponential, and no lower bound is known.

These tasks are those that reduce to polynomially many additive estimates of the quantities of Secs.~\ref{sec:ext}--\ref{sec:stats}: amplitudes and pattern weights of order one, traced-memory probabilities, overlaps and kernel entries, few-mode marginals and linear statistics. For them, the dequantisation known for Fock and Gaussian inputs \cite{Gurvits,AH,LimOh,Seron} extends to entangled inputs of polynomial bond dimension. Other tasks are not covered: relative estimation of exponentially small probabilities, and sampling from the full output distribution. Nor do we compare wall-clock times per device shot and per classical draw.

Three remarks help to read this bound correctly. First, $2^k$ is the memory available, not the memory used: a state produced by many emitters can still have a small bond dimension in emission order, and it is then easy as well. Second, the bound is an upper bound on the classical cost. A linear number of emitters is necessary for exponential hardness of these tasks but not sufficient, and our algorithm itself gives no lower bound; Sec.~\ref{sec:open} shows, however, that a cost polynomial in $k$ is excluded in general unless $\mathsf{BPP}=\mathsf{BQP}$. Third, architectures that add adaptive measurements, such as spin-optical schemes \cite{SPOQC}, can be universal, but they fall outside the fixed-interferometer setting considered here.

For photonic machine learning, the consequence is practical. A model fed by a few spins, whose loss is built from estimated probabilities, marginals or kernel entries, can be trained and evaluated classically, just as one fed by single photons. A route to an advantage through the source has to involve quantum memories that grow with the problem size, adaptive measurements, or a use of the device that does not reduce to additive estimates, such as sampling. Whether sampling from photonic MPS inputs is classically hard is a separate question, which we do not address here; exact classical sampling for these inputs is the subject of a companion paper \cite{ChainRules}.

\section{Numerical checks, cost and formalisation}
\label{sec:numbers}
\emph{Correctness.} We checked the identities and moments above against brute-force enumeration of~\eqref{eq:amp} with two independent NumPy implementations, and tested a PyTorch implementation of Algorithm~\ref{alg:sampler} against a permanent-sum oracle. The test inputs include projected and traced linear cluster, GHZ and random MPS states with $\chi=2$--$4$, collision-free and bunched outputs, and Haar-random interferometers. Over the 30 mean estimates of Table~\ref{tab:valid} (Supplementary Material, Sec.~\ref{app:numbers}), the largest deviation from the exact value is $2.4$ empirical standard errors, and the empirical second moments agree with their exact values within the fluctuations expected from Sec.~\ref{sec:tails}. A third implementation checks Corollary~\ref{cor:traced} and the factorised sampler of Remark~\ref{rem:factor} against exact moments and amplitudes on 36 states with $n$ up to $40$ and $\chi$ up to $32$ (Supplementary Material, Sec.~\ref{app:numbers}).

\emph{Cost.} Table~\ref{tab:h100} gives the cost per draw on one GPU. With Gaussian sampling it ranges from $4\,\mu$s ($n=10$, $\chi=2$) to $58\,\mu$s ($n=60$, $\chi=256$): $10^5$ draws, enough for an additive error of about $3\cdot10^{-3}$ on a normalised amplitude, take a few seconds even for a 60-photon input produced by eight two-level emitters. The dependence on $\chi$ is much milder than the $\chi^2$ of Theorem~\ref{thm:is} at these sizes, because the contractions are batched over draws. The sign (Rademacher) variant is up to $47$ times cheaper per draw for $\chi\le16$, partly because it avoids the eigenproblem (which Remark~\ref{rem:factor} also avoids for Gaussian sampling, but which Table~\ref{tab:h100} still uses), but Theorem~\ref{thm:is}(ii) does not hold for it: its second moment depends on the interferometer and can exceed $\|\psi\|^2$ (Sec.~\ref{sec:rademacher}). Its lower cost per draw is therefore not a lower cost at fixed accuracy.

\begin{table}[htbp]
\centering
\caption{Cost per draw on one NVIDIA H100 PCIe, amortised over $10^5$ draws (median of three runs), $m=2n$, Haar-random $U$, complex128. Gaussian / Rademacher sampling, in $\mu$s. $\chi=2$: projected linear cluster; $\chi=16,256$: random isometric emitters. Details in the Supplementary Material, Sec.~\ref{app:numbers}.}
\label{tab:h100}
\begin{tabular}{lccc}
\toprule
$n$ & $\chi=2$ & $\chi=16$ & $\chi=256$\\
\midrule
10 & 4.0 / 0.17 & 5.9 / 0.13 & 9.8 / 4.4\\
20 & 6.3 / 0.34 & 10.7 / 0.25 & 19.3 / 8.8\\
30 & 9.2 / 0.51 & 16.1 / 0.38 & 28.9 / 13.2\\
40 & 12.0 / 0.67 & 21.2 / 0.50 & 38.5 / 17.7\\
50 & 17.4 / 0.84 & 29.2 / 0.62 & 48.0 / 22.0\\
60 & 17.7 / 0.99 & 32.2 / 0.75 & 57.9 / 26.3\\
\bottomrule
\end{tabular}
\end{table}

\emph{Formalisation.} A Lean~4 project formalises Proposition~\ref{thm:identity}, parts (i) and (ii) of Theorem~\ref{thm:is}, the exactness of the sequential sampler including the local Gaussian draw of Algorithm~\ref{alg:sampler} (not the factorised draw of Remark~\ref{rem:factor}), and the vacuum rejection step (Proposition~\ref{prop:vacuum}). The operation counts, the numerical results and the complexity discussion are not formalised; Sec.~\ref{app:lean} maps each formal statement to its source file.

\section{Conclusion and open questions}
\label{sec:open}
Boson samplers are believed to be hard to simulate as samplers, but most of their proposed applications do not use samples directly. They rely on probabilities, marginals, overlaps or kernel entries estimated to additive precision, and for product inputs Gurvits's algorithm and its coherent-state form by Lim and Oh make these estimates classically easy. We have shown that the same holds for any input that is a matrix-product state with one photon per site and polynomial bond dimension. The estimator is unbiased, its second moment is the squared norm of the input whatever the interferometer, the output pattern or the amount of entanglement, and its cost grows polynomially with $\chi$. The same sampler gives probabilities with the memory traced out, overlaps and kernel entries between two such states, and individual marginal probabilities.

Since quantum emitters produce exactly such states, with $\chi$ set by the size of their memory, entangled light from a few spins does not by itself make estimation-based applications of linear optics hard to simulate. An advantage for these tasks has to come from a quantum memory that grows with the number of photons, from adaptive measurements, or from uses of the device that do not reduce to additive estimates. For these tasks, what matters is therefore not the entanglement of the input but the size of the memory that produces it: a linear cluster state from one spin is as easy to estimate as single photons, despite its large multipartite entanglement. Our estimator gives an upper bound on the classical cost, $\operatorname{poly}(n)\,2^{O(k)}$ for $k$ two-level emitters. A conditional lower bound shows that this growth with $k$ cannot be removed in general. When the emitters interact, their memory can run an arbitrary $k$-qubit circuit $C$ between emissions, and a projected weight such as $|\bra f M^{(1)}_{x_1}C\ket{s_0}|^2$ can then encode the acceptance probability of a polynomial-size quantum computation. Estimating it to inverse-polynomial additive error is $\mathsf{BQP}$-hard, so no classical algorithm can have a cost polynomial in $k$ for all such sources unless $\mathsf{BPP}=\mathsf{BQP}$. The true worst-case cost is therefore superpolynomial in $k$ and, by our algorithm, at most exponential; this is why the threshold $k=O(\log n)$ is meaningful rather than an artefact of our method. The argument uses interacting emitters; for more restricted sources, such as non-interacting spins with fixed rotations, the cost could be lower. Spin emitters combined with linear optics, photon detection and feed-forward also form a universal architecture \cite{SPOQC}. An experiment that aims at an advantage in estimation tasks with emitter-generated light needs a memory of more than $O(\log n)$ qubits, or a use of the output that does not reduce to additive estimates.

Several questions remain open. Sites holding several photons can already be included when their local Fock support is small and explicitly listed, with a rejection sampler whose mean cost is the number of local configurations (Sec.~\ref{app:local-fock}), but it would be useful to avoid this factor when the local support is large. Expectation values of general product observables need their own representation and variance bound, as Lim and Oh provide for product inputs \cite{LimOh}. Beyond matrix-product states, tree tensor networks, which describe some multi-emitter graph states, are natural candidates for the same continuous perfect-sampling approach. Real emitters also bring imperfections that a pure, fixed-photon-number MPS does not describe, such as partial distinguishability, multi-photon emission and unresolved loss, and it would be valuable to know how far the estimator extends to them.

Finally, whether sampling from the output distribution is classically hard for MPS inputs, including with a traced memory, remains open. Additive estimates alone do not yield an efficient sampler without further structure \cite{PBG}, and exact sampling for these inputs is the subject of a companion paper \cite{ChainRules}.

\section*{Reproducibility} The code is the MIT-licensed package \texttt{gurvits-mps}, in the folder \path{code/} of the archive of this work, \url{https://doi.org/10.5281/zenodo.24023358}, and the Lean proofs are in its folder \path{Lean/} (Supplementary Material, Sec.~\ref{app:lean}). The package contains the estimator (\path{gurvits_mps/estimate.py}: Algorithm~\ref{alg:sampler}, Gaussian and Rademacher sampling, variants A/B, traced memory by Corollary~\ref{cor:traced} or a Gaussian final vector, overlaps; PyTorch, CPU or GPU), a brute-force oracle (\path{gurvits_mps/reference.py}), tests, and the scripts producing every number of this work: \path{check_identity.py} (Table~\ref{tab:valid}, NumPy only, seed 1, about three minutes on one core), \path{validate.py} (Sec.~\ref{app:numbers}), \path{timing.py} (Table~\ref{tab:h100}), \path{haar_mean_scan.py} (\eqref{eq:haarmean} and the DFT$_4$ case), \path{check_vacuum.py} (Proposition~\ref{prop:vacuum}) and \path{traced_factorised/} (Corollary~\ref{cor:traced} and Remark~\ref{rem:factor}, NumPy and SciPy), all in \path{scripts/}.

\section*{Acknowledgements}
We thank Rawad Mezher for his insightful feedback and inspiration.
This work has been co-funded by the European Commission by the Horizon-CL4 program under the grant agreement 101135288 for the EPIQUE project, by the PROQCIMA and TUF-TOPIQC programs within the French National Quantum Strategy (France 2030).
The package code, the Lean proofs (Supplementary Material, Sec.~\ref{app:lean}), and the manuscript were written with the assistance of generative models from Anthropic (Claude). OpenAI (ChatGPT) assisted with the revision of the manuscript. The algorithms, the formalisation choices, the code architecture, the main definitions, the theorems and the propositions were designed and written or reviewed by the authors, who take responsibility for the whole.

\clearpage
\appendix
\section*{Supplementary material}

\setcounter{section}{0}
\setcounter{equation}{0}
\setcounter{table}{0}

\renewcommand{\thesection}{S\arabic{section}}
\renewcommand{\theequation}{S\arabic{equation}}
\renewcommand{\thetable}{S\arabic{table}}

\renewcommand{\theHsection}{supp.\arabic{section}}
\renewcommand{\theHequation}{supp.\arabic{equation}}
\renewcommand{\theHtable}{supp.\arabic{table}}

\setcounter{theorem}{0}
\setcounter{proposition}{0}
\setcounter{corollary}{0}
\setcounter{remark}{0}
\setcounter{conjecture}{0}

\renewcommand{\thetheorem}{S\arabic{theorem}}
\renewcommand{\theproposition}{S\arabic{proposition}}
\renewcommand{\thecorollary}{S\arabic{corollary}}
\renewcommand{\theremark}{S\arabic{remark}}
\renewcommand{\theconjecture}{S\arabic{conjecture}}

\renewcommand{\theHtheorem}{supp.\arabic{theorem}}
\renewcommand{\theHproposition}{supp.\arabic{proposition}}
\renewcommand{\theHcorollary}{supp.\arabic{corollary}}
\renewcommand{\theHremark}{supp.\arabic{remark}}
\renewcommand{\theHconjecture}{supp.\arabic{conjecture}}

The main text gives the estimator, its applications and its statistical interpretation. The following sections supply extensions of the input model and technical details. Sections~\ref{app:vacuum}--\ref{app:loss} treat vacuum emissions, finite local Fock support and known loss. Section~\ref{sec:rademacher} analyses sign sampling, and Sec.~\ref{app:tails} explains the heavy tails. Section~\ref{app:remarks} discusses related simulation tasks. Detailed numerical checks and the Lean correspondence appear in Secs.~\ref{app:numbers} and~\ref{app:lean}.

\section{Vacuum emissions}
\label{app:vacuum}

Emissions that may be vacuum fit the scheme of the main text. Let there be $N_{\mathrm e}$ emissions, each with label $x_k\in\{0,1,\dots,L\}$, where $x_k=0$ is the vacuum (operator $M^{(k)}_0$, no photon) and $x_k=l\ge1$ puts one photon in mode $\mu_{k,l}$, so that $\ket\psi=\sum_\bx\psi(\bx)\ket\bx$ with $\psi(\bx)=\bra fM^{(N_{\mathrm e})}_{x_{N_{\mathrm e}}}\cdots M^{(1)}_{x_1}\ket{s_0}$ spans several photon numbers. Let $|\bx|$ be the number of nonzero labels of $\bx$, and $c(\bx)$ the list of the modes $\mu_{k,x_k}$ with $x_k\ne0$. For an output $\bt$ with $n\le N_{\mathrm e}$ photons, $A_\bt=(\bt!)^{-1/2}\sum_{|\bx|=n}\psi(\bx)\per(U[r,c(\bx)])$, the sum running over the words with $n$ photons, and $\Phi(x)=\bra f\prod_k\big(M^{(k)}_0+\sum_{l\ge1}\bar x_{\mu_{k,l}}M^{(k)}_l\big)\ket{s_0}$.

The extra label changes the sampler in one place. The next-site amplitude is now a constant plus a linear function of the Gaussian vector. We first sample an incoherent mixture of these two contributions, then use rejection to restore their interference.

\begin{proposition}[Vacuum emissions]
\label{prop:vacuum}
In this setting the Gaussian statement of Proposition \mref{thm:identity} and parts (i)--(ii) of Theorem \mref{thm:is} hold with $\|\psi\|^2=\sum_\bx|\psi(\bx)|^2$ summed over all photon numbers. The density $q$ can be sampled exactly, emission by emission, by a rejection step whose acceptance probability is exactly $1/2$ at every emission; the expected cost per draw is $O(N_{\mathrm e}L\chi^2+N_{\mathrm e}L^3+mn)$ after $O(N_{\mathrm e}L\chi^3)$ preprocessing.
\end{proposition}
\begin{proof}
For independent standard complex Gaussians the moment $\E[\overline{x^T}x^c]$ factorises over the modes and vanishes unless every mode has the same exponent in both monomials; in particular it vanishes when the degrees differ. The degree-$n$ polynomial $g_\bt$ therefore pairs only with the words of $n$ photons, to which the proof of Proposition \mref{thm:identity} applies unchanged. The vacuum label uses no mode, so distinct words still have distinct mode sets, and the proofs of parts (i) and (ii) go through with the sum over all words. 

\paragraph{The conditional density.}
For sampling, set $R_{k-1}=\sum_{l=0}^LM^{(k)\dagger}_lR_kM^{(k)}_l$ and, given $a_{k-1}$, $b_l=M^{(k)}_la_{k-1}$ ($l=0,\dots,L$). As in the proof of Theorem \mref{thm:is}(iii), the conditional density of $v_k$ is
\[
  \gamma(v)\,\frac{\|h+Bv\|^2}{\|h\|^2+\Tr G_k},\qquad \|h+Bv\|^2=a_k^\dagger R_ka_k,\quad \|h\|^2=b_0^\dagger R_kb_0,\quad \|Bv\|^2=v^\dagger G_kv,
\]
with $v=v_k=(v_l)_{l\ge1}$, $a_k=b_0+\sum_{l\ge1}v_lb_l$, $h=R_k^{1/2}b_0$, the columns of $B$ equal to $R_k^{1/2}b_l$, and $G_k$ as in \meqref{eq:cond}; the denominator follows from $\E_\gamma v=0$ and $\E_\gamma vv^\dagger=\id$. 

\paragraph{A proposal that is easy to draw.}
This is no longer a homogeneous quadratic form, but $\|h+Bv\|^2\le2(\|h\|^2+\|Bv\|^2)$. Propose $v$ from $\pi(v)=\gamma(v)(\|h\|^2+v^\dagger G_kv)/(\|h\|^2+\Tr G_k)$, a mixture of $\gamma$ (weight $\|h\|^2/(\|h\|^2+\Tr G_k)$) and of the density \meqref{eq:cond} drawn as in Algorithm \mref{alg:sampler}, and accept it with probability $\|h+Bv\|^2/\big(2(\|h\|^2+v^\dagger G_kv)\big)\le1$. 

\paragraph{Why rejection gives the right law.}
An accepted $v$ has density proportional to $\pi(v)$ times this probability, that is to $\gamma(v)\|h+Bv\|^2$, and the acceptance probability is $\int\gamma\|h+Bv\|^2/\big(2(\|h\|^2+\Tr G_k)\big)=1/2$. The number of proposals per emission is geometric with mean $2$, and each costs $O(L^2\chi+L)$ once the $b_l$ and $R_kb_l$ are formed. No square root of $R_k$ is needed.
\end{proof}
The Rademacher identity is not asserted for photon-number superpositions: real signs do not distinguish all monomial degrees. The second moment is bounded by the norm of the whole train, not of its $n$-photon sector, so the bound is loose when the $n$-photon sector carries little weight. An independent check (\path{check_vacuum.py}, Sec.~\ref{app:numbers}) confirms unbiasedness, the second moment and the acceptance probability $1/2$.

\section{Finite local Fock support}
\label{app:local-fock}
The same idea gives a simple extension when an emission block can contain several photons. Suppose its allowed Fock configurations form an explicitly listed set $\mathcal A$ of size $D$, on modes disjoint from other blocks. For $\alpha\in\mathcal A$, use the Gaussian monomial
\[
 h_\alpha(v)=\prod_j\frac{v_j^{\alpha_j}}{\sqrt{\alpha_j!}}.
\]
These monomials are orthonormal under the product complex Gaussian measure. With $b_\alpha=M_\alpha a$ and $w_\alpha=R^{1/2}b_\alpha$, the next conditional density is
\[
 \frac{\gamma(v)}{Z}\left\|\sum_{\alpha\in\mathcal A}h_\alpha(v)w_\alpha\right\|^2,
 \qquad Z=\sum_\alpha\|w_\alpha\|^2.
\]
There is no need to solve a general polynomial sampling problem. Propose from
\[
 \pi(v)=\frac{\gamma(v)}{Z}\sum_\alpha|h_\alpha(v)|^2\|w_\alpha\|^2.
\]
Choose $\alpha$ with weight $\|w_\alpha\|^2/Z$; independently for each mode draw a uniform phase and squared radius $\mathrm{Gamma}(\alpha_j+1,1)$. This is exactly the selected mixture component. By Cauchy--Schwarz,
\[
 \left\|\sum_\alpha h_\alpha(v)w_\alpha\right\|^2
 \le D\sum_\alpha |h_\alpha(v)|^2\|w_\alpha\|^2.
\]
Accept with the ratio of the left-hand side to the right-hand side, taking any value on the proposal-null set where both vanish. Proposal density times acceptance probability is the target density divided by $D$. Acceptance therefore has probability $1/D$, and the mean number of proposals is $D$. All norms and inner products can be evaluated with $R$, without forming its square root.

The corresponding input polynomial is $\Phi(x)=\sum_\bs\psi_\bs\bar x^{\bs}/\sqrt{\bs!}$. Gaussian pairing gives $\E_\gamma[\Phi g_\bt]=\sqrt{\bt!}A_\bt$, and its importance estimator has second moment $\|\psi\|^2$ after output normalisation, as before. Thus an explicitly given, polynomial-size local support admits a polynomial expected-cost extension, counting the local photon occupations among the input parameters. This argument does not make an exponentially large local configuration list efficient. The vacuum sampler is the sharper degree-one construction; it has mean proposal count two, independently of $L$.

\section{Unused modes and known loss}
\label{app:loss}
\begin{remark}[Unused modes]
\label{rem:zero}
Only the unitarity of $U$ enters part (ii), through $\E_\gamma|g_\bt|^2=\per\big((UU^\dagger)[r,r]\big)$, which is Wick's theorem for the Gaussian vector $y=Ux$ of covariance $UU^\dagger$. (A) With $x$ standard Gaussian on all $m$ modes (Algorithm \ref{alg:sampler} as written) and $U$ unitary, this is $\bt!$. (B) If instead $x$ is set to zero on the $m-Ln$ modes the state does not use ($S$ the set of modes used by the state, $\bar S$ its complement, $U_S$ and $U_{\bar S}$ the corresponding columns of $U$), the identity still holds (those monomials of $g_\bt$ had zero expectation against $\Phi$), the sampling measure is different, $g_\bt$ costs $O(Ln^2)$ instead of $O(mn)$, and the second moment becomes $\|\psi\|^2\per\big((U_SU_S^\dagger)[r,r]\big)\le\|\psi\|^2\bt!$, by superadditivity of the permanent on positive-semidefinite matrices (Marcus--Newman \cite{MarcusNewman}, or the Cauchy--Binet expansion of $\per(KK^\dagger)$, for any matrix $K$, as a sum of non-negative terms) since $U_SU_S^\dagger+U_{\bar S}U_{\bar S}^\dagger=\id$ and $\per(\id[r,r])=\bt!$. At $m=Ln$ the two variants coincide; when $m\gg Ln$, variant B reduces the second moment without increasing the evaluation cost (Table \ref{tab:valid}, last projected row: $0.77$ against $2.87$ at $m=3n$).
\end{remark}

\subsection{Known loss}

Uniform or mode-dependent transmission that is known in advance is folded into the interferometer, which becomes a sub-unitary matrix $U$ ($UU^\dagger\preceq\id$), and we estimate the joint weight for a specified $n$-photon output and no lost photons, for a fixed $n$-photon input. A probability conditional on survival requires division by the survival probability; the unconditional second-moment bound below must then be rescaled. For an input spanning several photon numbers, a sub-unitary describes its no-loss branch only, not contributions from higher-number sectors that lose photons. Proposition \mref{thm:identity} holds for every $m\times m$ matrix $U$, so the identity and the unbiasedness in part (ii) of Theorem \mref{thm:is} are unchanged. Only the second moment uses unitarity: by Wick's theorem for the Gaussian vector $Ux$, of covariance $UU^\dagger$, $\E_\gamma|g_\bt|^2=\per\big((UU^\dagger)[r,r]\big)$. The matrices $(UU^\dagger)[r,r]$ and $(\id-UU^\dagger)[r,r]$ are positive semidefinite, and the permanent is superadditive on such matrices (Remark \mref{rem:zero}), so $\per\big((UU^\dagger)[r,r]\big)\le\per(\id[r,r])=\bt!$ and $\E_q|Z|^2\le\|\psi\|^2\,\bt!$. Unresolved loss, with a trace over the lost photons and several photon-number sectors, is not covered.

\section{Rademacher variant}
\label{sec:rademacher}
Let $\mu$ be the uniform measure on $\{\pm1\}^m$, with fixed photon number and distinct occupied input modes as in the main model. Orthogonality of the sign monomials gives $\E_\mu|\Phi|^2=\|\psi\|^2$. Thus $q_R(x)=\mu(x)|\Phi(x)|^2/\|\psi\|^2$ is a probability mass function and $Z=\|\psi\|^2g_\bt/\bar\Phi$ remains unbiased. Each conditional is proportional to $v^\dagger G_kv$ on $\{\pm1\}^L$. Enumerating these $2^L$ vectors avoids an eigenproblem and is useful for small fixed $L$; it is not polynomial in unrestricted $L$.

There is an important difference from Gaussian sampling. A nonzero polynomial can vanish on sign vectors with positive probability. Cancellation therefore gives
\[
 \E_{q_R}|Z|^2
 =\|\psi\|^2\E_\mu\!\left[|g_\bt(x)|^2\mathbf1_{\{\Phi(x)\ne0\}}\right]
 \le\|\psi\|^2\E_\mu|g_\bt(x)|^2.
\]
The equality without the indicator holds if $\Phi$ is nonzero on every sign vector, as for a single Fock word. It is not valid for arbitrary superpositions. For example, take one photon in $(\ket{10}+\ket{01})/\sqrt2$, a Hadamard interferometer, and the output row $(1,-1)/\sqrt2$. Then $\Phi=(x_1+x_2)/\sqrt2$ and $g=(x_1-x_2)/\sqrt2$. On the support of $q_R$, $g=0$, so the second moment is zero although $\E_\mu|g|^2=1$.

Even the upper bound depends on the interferometer. For the unitary $\mathrm{DFT}_4$ with rows numbered $0,1,2,3$, rows $(1,3)$ give $\E_\mu|g|^2=3/2$, while rows $(0,1)$ give $1/2$. The former exceeds $\bt!=1$. A general pointwise bound is
\[
 |g_\bt(x)|^2\le(t_{\max}m/n)^n,\qquad t_{\max}=\max_jt_j,
\]
by arithmetic--geometric mean and $\|Ux\|^2=m$. This supplies a uniform bound, not a claim that it is the best possible second-moment bound.

For Haar-random $U$, fixed $x$ is sent to a uniform vector on the sphere of squared radius $m$. For collision-free rows its coordinate moments give
\begin{equation}
 \E_U\E_\mu|g_\bt(x)|^2
 =\frac{m^n}{m(m+1)\cdots(m+n-1)}.
 \label{eq:haarmean}
\end{equation}
For a fixed input, the Haar-averaged second moment of $Z$ is this expression times $\|\psi\|^2\mu\{\Phi\ne0\}$. This follows because the Haar average is the same for every sign vector. For $\hat Z$, divide all second moments by $\bt!$. The timing advantage of signs in Table~\ref{tab:h100} concerns cost per draw; it does not establish a uniform advantage at fixed estimation accuracy.

\section{Heavy tails}
\label{app:tails}
$\hat Z=\|\psi\|^2g_\bt/(\sqrt{\bt!}\,\bar\Phi)$ has a finite second moment but, in general, an infinite fourth moment: near a simple zero of $\Phi$ where $g_\bt$ is nonzero, the density $q\propto\gamma|\Phi|^2$ vanishes quadratically while $|\hat Z|^4\propto|\Phi|^{-4}$, and $\int|\Phi|^{-2}$ diverges logarithmically in two real dimensions. The simplest instance is $n=1$, $m=2$, $\ket\psi=\ket1$, $U$ the Hadamard matrix, $\bt=(1,0)$: $\Phi=\bar x_1$, $\hat Z=(x_1+x_2)/(\sqrt2\,x_1)$, $\E_q|\hat Z|^2=1$ and $\E_q|\hat Z|^4=\infty$. In the reported numerical examples, the bulk of $|\hat Z|^2$ also lies far below its mean. Measured under $q$ on Haar-random $U$, the mean of $\log|\hat Z|^2$ decreases linearly in $n$ with slope $-0.58$, while $\E_q|\hat Z|^2$ stays fixed. A heuristic for the order of magnitude: under $\gamma$ and for collision-free $\bt$, $|g_\bt|^2$ is a product of $n$ i.i.d.\ $\mathrm{Exp}(1)$ variables, whose logarithm has mean $-\gamma_{\rm E}n$ ($\gamma_{\rm E}=0.5772$) although $\E|g_\bt|^2=1$; under $q$ the denominator $|\Phi|^2$ contributes as well, and we do not claim that the measured slope follows from the numerator alone. Consequently the Bessel-corrected sample variance, although unbiased, is extremely right-skewed, and a typical finite-run variance estimate can severely underestimate the true variance: at $N=10^5$ the empirical second moment of a $\chi=2$ projected cluster state fell from $0.08$ at $n=10$ to $2\cdot10^{-5}$ at $n=40$ against the exact value $\|\psi\|^2=\tfrac12$. We therefore report the theorem-based bound --- by Chebyshev, $|\hat A-A_\bt|\le\sqrt{\|\psi\|^2/(N\delta)}$ with probability $1-\delta$ --- and, for high-confidence estimates, a median-of-means estimator (real and imaginary parts separately) with $O(\|\psi\|^2\varepsilon^{-2}\log(1/\delta))$ draws for confidence $1-\delta$.

\section{Further remarks}
\label{app:remarks}
Whether sampling hardness can be paired with a classically trainable loss is a separate question: the few-mode marginals of Sec.~\ref{sec:marginals} are exactly the low-order statistics such a training would consume, and for generative models whose sampling hardness rests on anticoncentration, Herbst et al.\ \cite{Herbst} argue that such losses concentrate.

The identity needs a fixed input state. It does not cover adaptive protocols in which a mid-sequence measurement of the emitter conditions the later interferometer: a bounded number of adaptive steps is handled by enumerating outcomes, and estimation can become hard when the number of adaptive steps scales, with thresholds that differ for mean values and for sampling \cite{OhLim,ChabaudMarkhamSohbi,Monbroussou}. It also says nothing about pattern-level statistics over exponentially many patterns, nor about measurement-induced entanglement across emitters. On the simulation side, the tensor-network samplers of \cite{OhMPS,OhLoss,ParkOh} represent the \emph{output} of lossy or noisy circuits as an MPS along the modes and sample it; this note concerns an MPS \emph{input} to an arbitrary-depth interferometer and estimates rather than samples.

\section{Validation and cost per draw}
\label{app:numbers}

Table \ref{tab:valid} is the output of an independent NumPy implementation of Algorithm \mref{alg:sampler} (\path{check_identity.py}, seed 1): random and GHZ-type sequential MPSs, Haar-random $U$, collision-free and bunched outputs, $m=2n$ and one case with $m=3n$, exact values by brute-force enumeration of \meqref{eq:amp}. It checks Proposition \mref{thm:identity} under plain $\gamma$ sampling, Theorem \mref{thm:is} with Algorithm \mref{alg:sampler}, variants A and B of Remark \mref{rem:zero}, the Rademacher variant, the Gaussian-final-vector estimator for a traced memory (Sec.~\ref{sec:traced}), the DFT$_4$ counterexample and \eqref{eq:haarmean}. Over the 30 mean estimates the largest $|z|$ (empirical standard error) is $2.4$; the empirical second moments scatter around their exact values on both sides (within about $13\%$ for the projected Gaussian entries at $N=2$--$4\cdot10^4$), the variant-B moment at $m=3n$ equals $\per((U_SU_S^\dagger)[r,r])$ to $5\%$, and the traced empirical moments fluctuate around $\E_z\|\psi_z\|^4$; a finite-run second moment need not lie below its expectation.

\paragraph{Vacuum emissions.} \path{check_vacuum.py} (NumPy, seed 1) implements the sampler of Proposition \ref{prop:vacuum} on four random sequential states with a vacuum label, $N_{\mathrm e}=4$--$5$ emissions, $\chi=2$--$3$, $n=1$--$3$ detected photons (one bunched output), $m=2N_{\mathrm e}$, $4\cdot10^4$ draws each. Both the plain-$\gamma$ identity and the sampled estimator agree with brute-force enumeration (largest $|z|=1.55$), the empirical second moments are within $0.93$--$1.08$ of $\|\psi\|^2\bt!$, and the measured acceptance probability per emission is $0.500$--$0.501$.

\begin{table}[htbp]
\centering
\caption{Independent validation (NumPy implementation of Algorithm \mref{alg:sampler}, seed 1). $z=|\text{estimate}-\text{exact}|/\text{empirical standard error}$; $M_2$ = empirical second moment of $Z$ (Gaussian: exact value $\|\psi\|^2\bt!$; variant B: exact value $\|\psi\|^2\per((U_SU_S^\dagger)[r,r])$, equal to the former when $m=nL$); the ``$\gamma$'' column is the plain identity \meqref{eq:identity} sampled under $\gamma$. Traced rows (Gaussian final vector, Sec.~\ref{sec:traced}): $M_2$ of $W$ against $\E_z\|\psi_z\|^4$ (the random MPS is not normalised). Haar-random $U$, $m=2n$ unless stated, $N=4\cdot10^4$ ($2\cdot10^4$ for $\chi=4$; $3\cdot10^3$ traced).}
\label{tab:valid}
\footnotesize\setlength{\tabcolsep}{4pt}
\begin{tabular}{llccc|c|cc|cc|cc}
\toprule
input & register & $n$ & $\chi$ & $\bt!$ & $\gamma$ & \multicolumn{2}{c|}{Gaussian, variant A} & \multicolumn{2}{c|}{Gaussian, variant B} & \multicolumn{2}{c}{Rademacher}\\
 & & & & & $|z|$ & $|z|$ & $M_2$ / exact & $|z|$ & $M_2$ / exact & $|z|$ & $M_2$\\
\midrule
random MPS & proj. & 4 & 2 & 1 & 0.11 & 1.41 & 0.358 / 0.359 & 1.00 & 0.354 / 0.359 & 0.08 & 0.210\\
random MPS & proj. & 5 & 3 & 1 & 0.48 & 0.74 & 20.54 / 20.27 & 1.71 & 20.09 / 20.27 & 2.43 & 9.84\\
random MPS, bunched & proj. & 5 & 2 & 4 & 1.19 & 1.41 & 11.71 / 11.40 & 1.58 & 11.86 / 11.40 & 0.64 & 5.22\\
random MPS & proj. & 6 & 4 & 1 & 0.53 & 0.75 & 38.89 / 38.65 & 0.76 & 34.75 / 38.65 & 0.37 & 13.51\\
GHZ & proj. & 5 & 2 & 1 & 1.10 & 1.47 & 0.515 / 0.500 & 2.41 & 0.513 / 0.500 & 0.61 & 0.115\\
GHZ, bunched & proj. & 4 & 2 & 2 & 1.24 & 1.42 & 0.900 / 1.000 & 0.82 & 1.127 / 1.000 & 1.09 & 0.460\\
random MPS, $m=3n$ & proj. & 4 & 2 & 1 & 1.05 & 0.64 & 3.07 / 2.87 & 1.35 & 0.730 / 0.767 & 1.09 & 1.43\\
\midrule
random MPS & traced & 4 & 2 & 1 & --- & 1.24 & 645 / 717 & --- & --- & --- & ---\\
GHZ & traced & 4 & 2 & 1 & --- & 0.81 & 1.69 / 1.50 & --- & --- & --- & ---\\
\bottomrule
\end{tabular}
\end{table}

\paragraph{Reference implementation.} The PyTorch class \texttt{GurvitsEstimator} uses complex128 arithmetic and supports Gaussian and Rademacher sampling, variants A/B, projected memory, traced memory by the basis-state estimator of Corollary \ref{cor:traced} (the default) or a Gaussian final vector (Sec.~\ref{sec:traced}), and overlaps. It is tested against the brute-force permanent-sum oracle of \path{reference.py} on 26 cases (\path{tests/}): projected linear-cluster, GHZ and random emitters at $n=3$--$4$ with $L=2,3$ and $\chi=2$--$3$, collision-free and bunched outputs, $m=nL$ and $m=3n$; traced linear-cluster and random emitters with $\chi=2$--$4$, normalised or not, by both routes; an overlap between two random emitters; and the plain-$\gamma$ identity. Every estimate is within $5\sigma$, every Gaussian second moment of an amplitude within $25\%$ of its exact value at $3\cdot10^4$ draws, and every second moment of Corollary \ref{cor:traced} within $30\%$ of $(\Tr\rho_{\mathrm{mem}})^2$ at $2$--$3\cdot10^4$ draws. \path{scripts/validate.py} prints the corresponding table (19 estimates on 9 cases, largest $|z|=2.8$ at $2\cdot10^4$ draws, $2\cdot10^3$ for the Gaussian final vector, seed 1).

\paragraph{Traced memory and the factorised sampler.} The estimator of Corollary \ref{cor:traced} and the sampler of Remark \ref{rem:factor} were checked with a separate NumPy and SciPy implementation (\path{scripts/traced_factorised/}; seeds 2026, 7 and 3 for the small states, the large states and the timings), alongside Algorithm \mref{alg:sampler} and the Gaussian-final-vector route. The small-state set has 28 states (random isometric emitters, random non-isometric MPSs, linear cluster and GHZ states) with $n=3$--$6$, $L=2$--$4$ and $\chi=1$--$8$, each with three outputs (collision-free at $m=nL$, collision-free at $m=nL+2$, and one bunched pair at $m=nL+2$), and eight independent runs per case of $4\cdot10^4$ draws ($10^4$ for the Gaussian final vector).

For the law of the sampled variables we use one- and two-site moments, $\E_q[\bar v_{k,a}v_{k,b}]$ and $\E_q[|v_{k,a}|^2|v_{k',b}|^2]$, whose exact values follow from the reduced density matrices of $\psi$ and whose sampling distributions have finite variance. In each batch a random linear combination of each kind gives one $z$-score. Eight further states with $n=15$--$40$, $L$ up to $8$ and $\chi$ up to $32$, whose exact moments are obtained by MPS contraction (six batches of $4\cdot10^4$ draws each), bring the total to $544$ $z$-scores per sampler, with Kolmogorov--Smirnov $p$-values $0.16$ (Algorithm \mref{alg:sampler}) and $0.45$ (Remark \ref{rem:factor}) against $\mathcal N(0,1)$ and largest $|z|$ of $3.1$ and $3.0$.

For unbiasedness we use $N|\bar Y-\text{exact}|^2/\mathrm{Var}_{\text{exact}}$, whose expectation is $1$, with the exact variances $r_\psi-|A_\bt|^2$, $(\Tr\rho_{\mathrm{mem}})^2-p^2$ and $(\Tr\rho_{\mathrm{mem}})^2+\Tr\rho_{\mathrm{mem}}^2-p^2$. Its mean over $672$ runs is $1.02\pm0.04$ and $0.99\pm0.04$ for amplitudes with the two samplers, $1.02\pm0.05$ and $1.05\pm0.04$ for Corollary \ref{cor:traced} with the two samplers, and $1.00\pm0.04$ for the Gaussian final vector. Because of the heavy tails of Sec.~\mref{sec:tails} this statistic has infinite variance, so the bootstrap errors quoted are indicative only; its medians are $0.67$--$0.69$ for all five estimators.

On a two-core CPU (batched NumPy, preprocessing excluded), the factorised sampler at $L=2$ runs at $0.9$--$1.1$ times the speed of Algorithm \mref{alg:sampler} for $\chi=16$--$128$ at $n=20$ and for $n=20$--$160$ at $\chi=16$ ($1.4$ at $n=10$), and is $1.3$--$2.5\times$ faster for $\chi\le8$. At $n=10$, $\chi=8$ it is faster by $2.6\times$, $6\times$ and $19\times$ at $L=4$, $16$ and $64$. At $n=20$, $L=2$, a draw of Corollary \ref{cor:traced} is $1.2$--$1.4\times$ faster than a Gaussian-final-vector draw for $\chi\le8$, and $2.3\times$ and $4\times$ faster at $\chi=16$ and $64$; with all $\chi$ sets of environments built in advance, the preprocessing is recovered after $20$--$1100$ draws. The results are in \path{results/traced_factorised_20261022/}.

\paragraph{Cost.} Table \ref{tab:h100} is the output of \path{scripts/timing.py} on one NVIDIA H100 PCIe (80 GB; driver 580.126.20, PyTorch 2.14.0+cu130, Python 3.12, complex128): $10^5$ draws per configuration in one batch after a warm-up batch, time measured after device synchronisation and divided by the number of draws (amortised throughput per draw, not latency), median of three repeats whose spread is below $7\%$ in every entry; projected linear cluster for $\chi=2$, random isometric emitters for $\chi=16,256$; $m=2n$, Haar-random $U$, one random collision-free output; the $O(nL\chi^3)$ right-environment preprocessing is included in the timed call. Peak device memory is $<0.5$ GB for $\chi\le16$ and $3.1$ GB for $\chi=256$. The growth with $n$ is roughly linear although the $O(mn)$ term of $g_\bt$ is quadratic, a batching effect at these sizes; the table documents the measured cost, not the asymptotics. The Rademacher variant is $18$--$47\times$ cheaper at $\chi\le16$ (no eigenproblem, real signs) and about $2\times$ at $\chi=256$, where the $\chi^2$ contractions dominate. The raw CSV files are in the package (\path{results/h100_20261003/}).

\section{Formal proofs in Lean}
\label{app:lean}

Proposition \mref{thm:identity} (Gaussian and Rademacher cases), parts (i)--(ii) of Theorem \mref{thm:is}, the exactness statement of part (iii) together with the sitewise draw of Algorithm \mref{alg:sampler}, and the identity, normalisation and rejection step of Proposition \ref{prop:vacuum} are formalised in Lean~4 with the Mathlib library. The project, archived at \url{https://doi.org/10.5281/zenodo.24023358} (folder \path{Lean/}), includes an \texttt{Axioms.lean} file that prints the dependencies of the principal declarations. Its README reports a successful build with only the standard axioms (propositional extensionality, choice and quotient soundness). Reproducing that check requires the pinned Lean and Mathlib versions and a fresh \texttt{lake build}. The sign-support correction, the finite-local-Fock extension and the complexity discussion are not covered by the listed formalisation. Modes, emissions and labels are arbitrary finite types, amplitudes complex numbers, and the measures are Mathlib measures on $\C^m$; nothing depends on floating-point arithmetic. Table \ref{tab:lean} lists the correspondence.

\begin{table}[htbp]\centering\footnotesize
\setlength{\tabcolsep}{4pt}
\begin{tabular}{>{\raggedright\arraybackslash}p{0.17\linewidth}>{\raggedright\arraybackslash}p{0.22\linewidth}>{\raggedright\arraybackslash}p{0.53\linewidth}}\toprule
paper & Lean file & main declarations and content\\\midrule
Proposition \mref{thm:identity}, proof & \texttt{Identity.lean} & \texttt{Pairs}: $\int\overline{x^T}x^c\,d\mu=\#\{\sigma:T\circ\sigma=c\}$ for distinct modes $T$; \texttt{gaussian\_identity}: under it, $\sum_w\psi(w)\per(U[r,c(w)])=\int\Phi\,g_\bt\,d\mu$ for every matrix $U$, every state with at most one photon per mode and every rows $r$ (collisions included)\\
pairing of monomials & \texttt{Words.lean} & \texttt{pairings\_eq}: $\#\{\sigma:c'\circ\sigma=c\}=\prod_at_a!$ when the occupations agree, $0$ otherwise; \texttt{wick\_of\_moments}: one-mode moments $p!\,\delta_{pq}$ give the Wick rule on the product measure\\
\S\mref{sec:identity}, Gaussian & \texttt{ComplexGaussian.lean} & \texttt{stdGaussian}: density $\pi^{-1}e^{-|z|^2}$; \texttt{stdGaussian\_moment}: $\int\bar z^pz^q=p!\,\delta_{pq}$ (polar coordinates); \texttt{complexGaussian\_identity}: Proposition \mref{thm:identity}\\
Proposition \mref{thm:identity}, Rademacher & \texttt{Rademacher.lean} & \texttt{rademacher\_pairs}, \texttt{rademacher\_identity}: the same identity with $x$ uniform on $\{\pm1\}^m$\\
Theorem \mref{thm:is} (i)--(ii) & \texttt{Estimator.lean} & \texttt{normalisation}: $\int|\Phi|^2d\gamma=\|\psi\|^2$; \texttt{qMeasure\_isProbability}: (i); \texttt{estimator\_mean}: $\E_qZ=\sqrt{\bt!}A_\bt$; \texttt{gOut\_second\_moment}: $\E_\gamma|g_\bt|^2=\bt!$ for unitary $U$; \texttt{estimator\_second\_moment}: $\E_q|Z|^2=\|\psi\|^2\bt!$ when $\{\Phi=0\}$ is $\gamma$-null\\
Theorem \mref{thm:is} (ii), null set & \texttt{NullSet.lean} & \texttt{gaussian\_Phi\_null}: $\gamma\{\Phi=0\}=0$ for $\psi\ne0$ ($\Phi$ is affine in each $\bar x_a$, and a slice $\{x_a:A+\bar x_aB=0\}$ with $B\ne0$ is a point); \texttt{theorem2\_gaussian\_full}: parts (i)--(ii) for $\psi\ne0$ and unitary $U$, with no further hypothesis\\
Theorem \mref{thm:is} (iii), proof & \texttt{Sampler.lean} & \texttt{env\_posSemidef}: the right environments $R_k$ are positive semidefinite; \texttt{integral\_wt\_step}: $\int\gamma(v)\,(\sum_lv_lb_l)^\dagger R(\sum_lv_lb_l)\,dv=\sum_lb_l^\dagger Rb_l$; \texttt{cond\_eq\_gram}: the sitewise conditional is \meqref{eq:cond}; \texttt{integral\_cond}: it is a probability density; \texttt{law\_eq}: drawing $v_0,\dots,v_{k-1}$ in turn from these conditionals gives the prefix the density $a_k^\dagger R_ka_k/s_0^\dagger R_0s_0$ with respect to $\gamma$\\
Theorem \mref{thm:is} (iii) & \texttt{SamplerLaw.lean} & \texttt{Phi\_output}: $\Phi(x)=\bra fa_n$ at the output $x_{\mu_{k,l}}=\bar v_{k,l}$; \texttt{pw\_zero\_eq\_normSq}: $s_0^\dagger R_0s_0=\|\psi\|^2$; \texttt{sampler\_exact}: with the unused modes drawn from $\gamma$, the output of the sequential draw has law $q$\\
Algorithm \mref{alg:sampler}, step 4 & \texttt{Mixture.lean}, \texttt{Radial.lean} & \texttt{map\_unitary\_gaussianVector}: $\gamma$ on $\C^L$ is invariant under unitaries; \texttt{sitewise\_draw}: with $G_k=U\,\mathrm{diag}(\lambda)\,U^\dagger$, drawing $j$ with probability $\lambda_j/\Tr G_k$ and $v=Uc$, with $c_j$ of density $\gamma(c)|c|^2$ and the other $c_i$ standard Gaussian (\texttt{tilted\_eq\_pi}), gives $v$ the density \meqref{eq:cond}; \texttt{radial\_draw}: $\sqrt s\,e^{i\theta}$ with $s\sim\mathrm{Gamma}(2,1)$ and $\theta$ uniform has density $\gamma(c)|c|^2$\\
Proposition \ref{prop:vacuum} & \texttt{Vacuum.lean} & \texttt{vacuum\_identity}: components of another photon number than the output add nothing to $\E_\gamma[\Phi g_\bt]$; \texttt{vacuum\_normalisation}: $\int|\Phi|^2d\gamma=\sum_\bx|\psi(\bx)|^2$ over all photon numbers; \texttt{target\_le}: $\|h+Bv\|^2\le2(\|h\|^2+\|Bv\|^2)$; \texttt{rejection\_subdensity}: proposal density times acceptance probability is $\gamma\|h+Bv\|^2/(2Z)$; \texttt{rejection\_acceptance}: acceptance probability $1/2$\\\bottomrule
\end{tabular}
\caption{Lean declarations corresponding to the mathematical statements. All declarations belong to the namespace \texttt{EstimatorNote}.}\label{tab:lean}
\end{table}

\noindent\emph{Modelling choices.} For parts (i)--(ii) the state is an arbitrary function on label words, as in the note, since they use no MPS structure; for part (iii) it is the sequential MPS \meqref{eq:mps} with bond space of any finite dimension, and the drawn law is compared with $q$ as a measure on $\C^m$. Lean's convention $a/0=0$ sets $Z=0$ on $\{\Phi=0\}$, which is null for both $q$ and $\gamma$.

\noindent\emph{Not formalised.} The operation counts of part (iii) of Theorem \mref{thm:is}; in Proposition \ref{prop:vacuum}, the form of the sitewise conditional with a vacuum label, part (ii) for the vacuum train, and the cost; Corollary \ref{cor:traced}; Remark \ref{rem:factor}; \S\S\ref{sec:ext}--\mref{sec:stats}; the inequalities of Remark \mref{rem:zero}; all cost statements.

\clearpage

\bibliographystyle{unsrtnat}
\bibliography{references}

\begin{thebibliography}{52}
\providecommand{\natexlab}[1]{#1}
\providecommand{\url}[1]{\texttt{#1}}
\expandafter\ifx\csname urlstyle\endcsname\relax
  \providecommand{\doi}[1]{doi: #1}\else
  \providecommand{\doi}{doi: \begingroup \urlstyle{rm}\Url}\fi

\bibitem[Aaronson and Arkhipov(2013)]{AA}
S.~Aaronson and A.~Arkhipov.
\newblock {The Computational Complexity of Linear Optics}.
\newblock \emph{Theory Comput.}, 9\penalty0 (4):\penalty0 143--252, 2013.
\newblock \doi{10.4086/toc.2013.v009a004}.

\bibitem[Clifford and Clifford(2018)]{CliffordClifford}
P.~Clifford and R.~Clifford.
\newblock {The Classical Complexity of Boson Sampling}.
\newblock In \emph{Proceedings of the 29th Annual ACM-SIAM Symposium on
  Discrete Algorithms (SODA)}, pages 146--155. SIAM, 2018.
\newblock \doi{10.1137/1.9781611975031.10}.

\bibitem[Broome et~al.(2013)Broome, Fedrizzi, Rahimi-Keshari, Dove, Aaronson,
  Ralph, and White]{Broome}
M.~A. Broome, A.~Fedrizzi, S.~Rahimi-Keshari, J.~Dove, S.~Aaronson, T.~C.
  Ralph, and A.~G. White.
\newblock {Photonic Boson Sampling in a Tunable Circuit}.
\newblock \emph{Science}, 339\penalty0 (6121):\penalty0 794--798, 2013.
\newblock \doi{10.1126/science.1231440}.

\bibitem[Spring et~al.(2013)]{Spring}
J.~B. Spring et~al.
\newblock {Boson Sampling on a Photonic Chip}.
\newblock \emph{Science}, 339\penalty0 (6121):\penalty0 798--801, 2013.
\newblock \doi{10.1126/science.1231692}.

\bibitem[Wang et~al.(2019)]{Wang2019}
H.~Wang et~al.
\newblock {Boson Sampling with 20 Input Photons and a 60-Mode Interferometer in
  a $10^{14}$-Dimensional Hilbert Space}.
\newblock \emph{Phys. Rev. Lett.}, 123:\penalty0 250503, 2019.
\newblock \doi{10.1103/PhysRevLett.123.250503}.

\bibitem[Hamilton et~al.(2017)Hamilton, Kruse, Sansoni, Barkhofen, Silberhorn,
  and Jex]{Hamilton}
C.~S. Hamilton, R.~Kruse, L.~Sansoni, S.~Barkhofen, C.~Silberhorn, and I.~Jex.
\newblock {Gaussian Boson Sampling}.
\newblock \emph{Phys. Rev. Lett.}, 119:\penalty0 170501, 2017.
\newblock \doi{10.1103/PhysRevLett.119.170501}.

\bibitem[Zhong et~al.(2020)]{Zhong}
H.-S. Zhong et~al.
\newblock {Quantum Computational Advantage Using Photons}.
\newblock \emph{Science}, 370\penalty0 (6523):\penalty0 1460--1463, 2020.
\newblock \doi{10.1126/science.abe8770}.

\bibitem[Madsen et~al.(2022)]{Madsen}
L.~S. Madsen et~al.
\newblock {Quantum Computational Advantage with a Programmable Photonic
  Processor}.
\newblock \emph{Nature}, 606:\penalty0 75--81, 2022.
\newblock \doi{10.1038/s41586-022-04725-x}.

\bibitem[Gan et~al.(2022)Gan, Leykam, and Angelakis]{Gan}
B.~Y. Gan, D.~Leykam, and D.~G. Angelakis.
\newblock {Fock State-Enhanced Expressivity of Quantum Machine Learning
  Models}.
\newblock \emph{EPJ Quantum Technol.}, 9:\penalty0 16, 2022.
\newblock \doi{10.1140/epjqt/s40507-022-00135-0}.

\bibitem[Sedrakyan and Salavrakos(2024)]{Sedrakyan}
T.~Sedrakyan and A.~Salavrakos.
\newblock {Photonic Quantum Generative Adversarial Networks for Classical
  Data}.
\newblock arXiv:2405.06023, 2024.

\bibitem[Gottlieb et~al.(2026)Gottlieb, Faraji, Mezher, Ventura, Mansfield, and
  Salavrakos]{Gottlieb}
F.~Gottlieb, C.~Faraji, R.~Mezher, B.~Ventura, S.~Mansfield, and A.~Salavrakos.
\newblock {Efficient Training of Photonic Quantum Generative Models}.
\newblock arXiv:2603.08793, 2026.

\bibitem[Gurvits(2005)]{Gurvits}
L.~Gurvits.
\newblock {On the Complexity of Mixed Discriminants and Related Problems}.
\newblock In \emph{Mathematical Foundations of Computer Science 2005 (MFCS)},
  volume 3618 of \emph{Lecture Notes in Computer Science}, pages 447--458.
  Springer, 2005.

\bibitem[Aaronson and Hance(2014)]{AH}
S.~Aaronson and T.~Hance.
\newblock {Generalizing and Derandomizing Gurvits's Approximation Algorithm for
  the Permanent}.
\newblock \emph{Quantum Inf. Comput.}, 14\penalty0 (7\&8):\penalty0 541--559,
  2014.
\newblock \doi{10.26421/QIC14.7-8-1}.

\bibitem[Lim and Oh(2026)]{LimOh}
Y.~Lim and C.~Oh.
\newblock {Classical Algorithms for Estimating Expectation Values in Linear
  Optical Circuits}.
\newblock \emph{Phys. Rev. Lett.}, 137:\penalty0 120602, 2026.
\newblock \doi{10.1103/bl9r-2gyb}.

\bibitem[Seron et~al.(2026)Seron, Thomas, Araujo, Arkhipov, Oh, and
  Novo]{Seron}
B.~Seron, H.~Thomas, E.~Araujo, A.~Arkhipov, C.~Oh, and L.~Novo.
\newblock {Efficient Classical Algorithm for Estimating Linear Statistics of
  Boson Sampling}.
\newblock arXiv:2609.35447, 2026.

\bibitem[Recio-Armengol et~al.(2025)Recio-Armengol, Ahmed, and
  Bowles]{RecioArmengol}
E.~Recio-Armengol, S.~Ahmed, and J.~Bowles.
\newblock {Train on Classical, Deploy on Quantum: Scaling Generative Quantum
  Machine Learning to a Thousand Qubits}.
\newblock arXiv:2503.02934, 2025.

\bibitem[Monbroussou et~al.(2026)Monbroussou, Thomas, Mhiri, Holmes, and
  Kashefi]{MonbroussouConc}
L.~Monbroussou, H.~Thomas, H.~Mhiri, Z.~Holmes, and E.~Kashefi.
\newblock {Classical Simulation and Model Concentration in Passive Linear
  Optics}.
\newblock arXiv:2607.24728, 2026.

\bibitem[Spagnolo et~al.(2023)Spagnolo, Brod, Galv{\~a}o, and
  Sciarrino]{Spagnolo}
N.~Spagnolo, D.~J. Brod, E.~F. Galv{\~a}o, and F.~Sciarrino.
\newblock {Non-Linear Boson Sampling}.
\newblock \emph{npj Quantum Inf.}, 9:\penalty0 3, 2023.
\newblock \doi{10.1038/s41534-023-00676-x}.

\bibitem[{Le Jeannic} et~al.(2022)]{LeJeannic}
H.~{Le Jeannic} et~al.
\newblock {Dynamical Photon--Photon Interaction Mediated by a Quantum Emitter}.
\newblock \emph{Nat. Phys.}, 2022.
\newblock \doi{10.1038/s41567-022-01720-x}.

\bibitem[Chabaud et~al.(2021)Chabaud, Markham, and Sohbi]{ChabaudMarkhamSohbi}
U.~Chabaud, D.~Markham, and A.~Sohbi.
\newblock {Quantum Machine Learning with Adaptive Linear Optics}.
\newblock \emph{Quantum}, 5:\penalty0 496, 2021.
\newblock \doi{10.22331/q-2021-07-05-496}.

\bibitem[Monbroussou et~al.(2025)Monbroussou, Mamon, Thomas, Yacoub, Chabaud,
  and Kashefi]{Monbroussou}
L.~Monbroussou, E.~Z. Mamon, H.~Thomas, V.~Yacoub, U.~Chabaud, and E.~Kashefi.
\newblock {Towards Quantum Advantage with Photonic State Injection}.
\newblock \emph{Phys. Rev. Res.}, 7:\penalty0 033051, 2025.
\newblock \doi{10.1103/PhysRevResearch.7.033051}.

\bibitem[Lindner and Rudolph(2009)]{LindnerRudolph}
N.~H. Lindner and T.~Rudolph.
\newblock {Proposal for Pulsed On-Demand Sources of Photonic Cluster State
  Strings}.
\newblock \emph{Phys. Rev. Lett.}, 103:\penalty0 113602, 2009.
\newblock \doi{10.1103/PhysRevLett.103.113602}.

\bibitem[Cogan et~al.(2023)Cogan, Su, Kenneth, and Gershoni]{Cogan}
D.~Cogan, Z.-E. Su, O.~Kenneth, and D.~Gershoni.
\newblock {Deterministic Generation of Indistinguishable Photons in a Cluster
  State}.
\newblock \emph{Nat. Photon.}, 17\penalty0 (4):\penalty0 324--329, 2023.
\newblock \doi{10.1038/s41566-022-01152-2}.

\bibitem[Thomas et~al.(2022)Thomas, Ruscio, Morin, and Rempe]{Thomas}
P.~Thomas, L.~Ruscio, O.~Morin, and G.~Rempe.
\newblock {Efficient Generation of Entangled Multiphoton Graph States from a
  Single Atom}.
\newblock \emph{Nature}, 608:\penalty0 677--681, 2022.
\newblock \doi{10.1038/s41586-022-04987-5}.

\bibitem[Coste et~al.(2023)]{Coste}
N.~Coste et~al.
\newblock {High-Rate Entanglement between a Semiconductor Spin and
  Indistinguishable Photons}.
\newblock \emph{Nat. Photon.}, 17\penalty0 (7):\penalty0 582--587, 2023.
\newblock \doi{10.1038/s41566-023-01186-0}.

\bibitem[Huet et~al.(2025)]{Huet}
H.~Huet et~al.
\newblock {Deterministic and Reconfigurable Graph State Generation with a
  Single Solid-State Quantum Emitter}.
\newblock \emph{Nat. Commun.}, 16:\penalty0 4337, 2025.
\newblock \doi{10.1038/s41467-025-59693-3}.

\bibitem[de~Gliniasty et~al.(2024)de~Gliniasty, Hilaire, Emeriau, Wein,
  Salavrakos, and Mansfield]{SPOQC}
G.~de~Gliniasty, P.~Hilaire, P.-E. Emeriau, S.~C. Wein, A.~Salavrakos, and
  S.~Mansfield.
\newblock {A Spin-Optical Quantum Computing Architecture}.
\newblock \emph{Quantum}, 8:\penalty0 1423, 2024.
\newblock \doi{10.22331/q-2024-07-24-1423}.

\bibitem[Sch{\"o}n et~al.(2005)Sch{\"o}n, Solano, Verstraete, Cirac, and
  Wolf]{Schon}
C.~Sch{\"o}n, E.~Solano, F.~Verstraete, J.~I. Cirac, and M.~M. Wolf.
\newblock {Sequential Generation of Entangled Multiqubit States}.
\newblock \emph{Phys. Rev. Lett.}, 95:\penalty0 110503, 2005.
\newblock \doi{10.1103/PhysRevLett.95.110503}.

\bibitem[Ferris and Vidal(2012)]{FerrisVidal}
A.~J. Ferris and G.~Vidal.
\newblock {Perfect Sampling with Unitary Tensor Networks}.
\newblock \emph{Phys. Rev. B}, 85:\penalty0 165146, 2012.
\newblock \doi{10.1103/PhysRevB.85.165146}.

\bibitem[Meiburg et~al.(2025)Meiburg, Chen, Miller, Tihon, Rabusseau, and
  Perdomo-Ortiz]{Meiburg}
A.~Meiburg, J.~Chen, J.~Miller, R.~Tihon, G.~Rabusseau, and A.~Perdomo-Ortiz.
\newblock {Generative Learning of Continuous Data by Tensor Networks}.
\newblock \emph{SciPost Phys.}, 18:\penalty0 096, 2025.
\newblock \doi{10.21468/SciPostPhys.18.3.096}.

\bibitem[Valiant(1979)]{Valiant}
L.~G. Valiant.
\newblock {The Complexity of Computing the Permanent}.
\newblock \emph{Theor. Comput. Sci.}, 8\penalty0 (2):\penalty0 189--201, 1979.
\newblock \doi{10.1016/0304-3975(79)90044-6}.

\bibitem[Lim and Oh(2023)]{LimOh2023}
Y.~Lim and C.~Oh.
\newblock {Approximating Outcome Probabilities of Linear Optical Circuits}.
\newblock \emph{npj Quantum Inf.}, 9:\penalty0 124, 2023.
\newblock \doi{10.1038/s41534-023-00791-9}.

\bibitem[{Correa Anguita} et~al.(2025){Correa Anguita}, Roelink, Marzban,
  Briels, Filippi, and Renema]{CorreaAnguita}
M.~{Correa Anguita}, T.~Roelink, S.~Marzban, W.~Briels, C.~Filippi, and
  J.~Renema.
\newblock {Experimental Demonstration of Boson Sampling as a Hardware
  Accelerator for Monte Carlo Integration}.
\newblock arXiv:2509.25404, 2025.

\bibitem[Oh et~al.(2021)Oh, Noh, Fefferman, and Jiang]{OhLoss}
C.~Oh, K.~Noh, B.~Fefferman, and L.~Jiang.
\newblock {Classical Simulation of Lossy Boson Sampling Using Matrix Product
  Operators}.
\newblock \emph{Phys. Rev. A}, 104:\penalty0 022407, 2021.
\newblock \doi{10.1103/PhysRevA.104.022407}.

\bibitem[Oh et~al.(2024)Oh, Liu, Alexeev, Fefferman, and Jiang]{OhMPS}
C.~Oh, M.~Liu, Y.~Alexeev, B.~Fefferman, and L.~Jiang.
\newblock {Classical Algorithm for Simulating Experimental Gaussian Boson
  Sampling}.
\newblock \emph{Nat. Phys.}, 20:\penalty0 1461, 2024.
\newblock \doi{10.1038/s41567-024-02535-8}.

\bibitem[Park and Oh(2025)]{ParkOh}
S.~Park and C.~Oh.
\newblock {Matrix Product State Approach to Lossy Boson Sampling and Noisy IQP
  Sampling}.
\newblock arXiv:2510.24137, 2025.

\bibitem[Schollw{\"o}ck(2011)]{Schollwock}
U.~Schollw{\"o}ck.
\newblock {The Density-Matrix Renormalization Group in the Age of Matrix
  Product States}.
\newblock \emph{Ann. Phys.}, 326\penalty0 (1):\penalty0 96--192, 2011.
\newblock \doi{10.1016/j.aop.2010.09.012}.

\bibitem[Vidal(2003)]{Vidal}
G.~Vidal.
\newblock {Efficient Classical Simulation of Slightly Entangled Quantum
  Computations}.
\newblock \emph{Phys. Rev. Lett.}, 91:\penalty0 147902, 2003.
\newblock \doi{10.1103/PhysRevLett.91.147902}.

\bibitem[Sch{\"o}n et~al.(2007)Sch{\"o}n, Hammerer, Wolf, Cirac, and
  Solano]{Schon2007}
C.~Sch{\"o}n, K.~Hammerer, M.~M. Wolf, J.~I. Cirac, and E.~Solano.
\newblock {Sequential Generation of Matrix-Product States in Cavity QED}.
\newblock \emph{Phys. Rev. A}, 75:\penalty0 032311, 2007.
\newblock \doi{10.1103/PhysRevA.75.032311}.

\bibitem[Economou et~al.(2010)Economou, Lindner, and Rudolph]{Economou2010}
S.~E. Economou, N.~Lindner, and T.~Rudolph.
\newblock {Optically Generated 2-Dimensional Photonic Cluster State from
  Coupled Quantum Dots}.
\newblock \emph{Phys. Rev. Lett.}, 105:\penalty0 093601, 2010.
\newblock \doi{10.1103/PhysRevLett.105.093601}.

\bibitem[Buterakos et~al.(2017)Buterakos, Barnes, and Economou]{Buterakos}
D.~Buterakos, E.~Barnes, and S.~E. Economou.
\newblock {Deterministic Generation of All-Photonic Quantum Repeaters from
  Solid-State Emitters}.
\newblock \emph{Phys. Rev. X}, 7:\penalty0 041023, 2017.
\newblock \doi{10.1103/PhysRevX.7.041023}.

\bibitem[Schwartz et~al.(2016)Schwartz, Cogan, Schmidgall, Don, Gantz, Kenneth,
  Lindner, and Gershoni]{Schwartz}
I.~Schwartz, D.~Cogan, E.~R. Schmidgall, Y.~Don, L.~Gantz, O.~Kenneth, N.~H.
  Lindner, and D.~Gershoni.
\newblock {Deterministic Generation of a Cluster State of Entangled Photons}.
\newblock \emph{Science}, 354\penalty0 (6311):\penalty0 434--437, 2016.
\newblock \doi{10.1126/science.aah4758}.

\bibitem[Istrati et~al.(2020)]{Istrati}
D.~Istrati et~al.
\newblock {Sequential Generation of Linear Cluster States from a Single Photon
  Emitter}.
\newblock \emph{Nat. Commun.}, 11, 2020.
\newblock \doi{10.1038/s41467-020-19341-4}.

\bibitem[Thomas et~al.(2024)Thomas, Ruscio, Morin, and Rempe]{Thomas2024}
P.~Thomas, L.~Ruscio, O.~Morin, and G.~Rempe.
\newblock {Fusion of Deterministically Generated Photonic Graph States}.
\newblock \emph{Nature}, 629\penalty0 (8012):\penalty0 567--572, 2024.
\newblock \doi{10.1038/s41586-024-07357-5}.

\bibitem[Besse et~al.(2020)]{Besse}
J.-C. Besse et~al.
\newblock {Realizing a Deterministic Source of Multipartite-Entangled Photonic
  Qubits}.
\newblock \emph{Nat. Commun.}, 11, 2020.
\newblock \doi{10.1038/s41467-020-18635-x}.

\bibitem[Ferreira et~al.(2024)Ferreira, Kim, Butler, Pichler, and
  Painter]{Ferreira}
V.~S. Ferreira, G.~Kim, A.~Butler, H.~Pichler, and O.~Painter.
\newblock {Deterministic Generation of Multidimensional Photonic Cluster States
  with a Single Quantum Emitter}.
\newblock \emph{Nat. Phys.}, 20\penalty0 (5):\penalty0 865--870, 2024.
\newblock \doi{10.1038/s41567-024-02408-0}.

\bibitem[Li et~al.(2022)Li, Economou, and Barnes]{LiEconomouBarnes}
B.~Li, S.~E. Economou, and E.~Barnes.
\newblock {Photonic Resource State Generation from a Minimal Number of Quantum
  Emitters}.
\newblock \emph{npj Quantum Inf.}, 8, 2022.
\newblock arXiv:2108.12466.

\bibitem[Senellart et~al.(2026)Senellart, Senellart, Wein, and
  Emeriau]{ChainRules}
J.~Senellart, P.~Senellart, S.~Wein, and P.-E. Emeriau.
\newblock {Exact Sampling beyond Fock Inputs and Linear Optics: Photon-Number
  Superpositions, Photonic Matrix-Product States, Adaptive and Non-Linear
  Circuits}.
\newblock In preparation, 2026.

\bibitem[Pashayan et~al.(2020)Pashayan, Bartlett, and Gross]{PBG}
H.~Pashayan, S.~D. Bartlett, and D.~Gross.
\newblock {From Estimation of Quantum Probabilities to Simulation of Quantum
  Circuits}.
\newblock \emph{Quantum}, 4:\penalty0 223, 2020.
\newblock \doi{10.22331/q-2020-01-13-223}.

\bibitem[Marcus and Newman(1962)]{MarcusNewman}
M.~Marcus and M.~Newman.
\newblock {Inequalities for the Permanent Function}.
\newblock \emph{Ann. Math.}, 75\penalty0 (1):\penalty0 47, 1962.

\bibitem[Herbst et~al.(2025)Herbst, Brandi{\'c}, and P{\'e}rez-Salinas]{Herbst}
A.~Herbst, I.~Brandi{\'c}, and A.~P{\'e}rez-Salinas.
\newblock {Limits of Quantum Generative Models with Classical Sampling
  Hardness}.
\newblock arXiv:2512.24801, 2025.

\bibitem[Oh and Lim(2025)]{OhLim}
C.~Oh and Y.~Lim.
\newblock {Classical Algorithms for Measurement-Adaptive Gaussian Circuits}.
\newblock arXiv:2509.00746, 2025.

\end{thebibliography}

\end{document}